\documentclass[11pt,oneside,letterpaper]{article}
\usepackage{amssymb}
\usepackage{amsmath}
\usepackage[dvips]{graphicx}
\usepackage{setspace}
\usepackage{amsfonts}
\usepackage{fancyhdr}
\usepackage{xcolor}
\usepackage{graphicx}
\usepackage{rotating}
\usepackage{comment}
\usepackage{color}
\usepackage{cite}
\usepackage{braket}
\usepackage{caption}
\usepackage[utf8]{inputenc}

\definecolor{darkgreen}{rgb}{0,0.5,0}
\definecolor{darkblue}{rgb}{0,0,0.6}
\definecolor{purple}{rgb}{0.4,.2,0.7}
\newcommand{\p}{\partial}

\newcommand{\f}{\frac}

\newcommand{\be}{\begin{equation}}
\newcommand{\ee}{\end{equation}}

\usepackage[colorlinks=true,citecolor=darkgreen,linkcolor=black,urlcolor=purple]{hyperref}

\usepackage{pdfsync}
\makeatletter
\newcommand*{\defeq}{\mathrel{\rlap{%
                     \raisebox{0.3ex}{$\m@th\cdot$}}%
                     \raisebox{-0.3ex}{$\m@th\cdot$}}%
                     =} 
\makeatother

\def\be{\begin{eqnarray}}
\def\ee{\end{eqnarray}}

\newcommand{\tr}{\textrm{Tr}\,}

\newcommand{\bea}{\begin{eqnarray}}
\newcommand{\eea}{\end{eqnarray}}
\def\ben{\begin{equation}}
\def\een{\end{equation}}

 \let\b=\beta   
   
\let\l=\lambda \let\m=\mu \let\n=\nu  \let\p=\phi \let\r=v
\let\s=\sigma

\let\f=\frac

\def\be{\begin{equation}}
\def\ee{\end{equation}}
\def\ba{\begin{array}}
\def\ea{\end{array}}

\def\ba#1\ea{\begin{align}#1\end{align}}
\def\bs#1\es{\begin{split}#1\end{split}}

\renewcommand{\p}{\partial}
\newcommand{\ie}{\textit{i.e.}}
\newcommand{\Ssemi}{S_{\rm mat}}
\newcommand{\Isemi}{I_{\rm mat}}
\newcommand{\epsilonuv}{\epsilon_\textup{uv}}
\newcommand{\Sdiv}{S_\textup{ct}}

\newcommand{\Sred}{\widehat{S}_{\rm mat}}
\newcommand{\Area}{\mbox{Area}}
\newcommand{\setcomp}{\mathsf{c}}
\newcommand{\bz}{\bar{z}}
\newcommand{\bsigma}{\bar{\sigma}}

\newcommand{\Sflat}{\widetilde{S}_{\rm mat}}
\newcommand{\Sfrw}{\Ssemi}
\newcommand{\tV}{\widetilde{V}}
\newcommand{\wArea}{\widetilde{\mbox{Area}}}

\newcommand{\tVol}{\widetilde{\mbox{Vol}}}
\newcommand{\epsilonrg}{\epsilon_{\rm rg}}

\allowdisplaybreaks  % allow page breaks in displayed eqs

\numberwithin{equation}{section}
\def \be {\begin{equation}}
\def \ee {\end{equation}}
	
\def \JM#1 {{\color{blue}  JM: #1 }}
\def \AAl#1 {{\color{red}  AA: #1 }}

\begin{document}
\onehalfspacing

\begin{center}

~
\vskip5mm

{\LARGE  {
Islands in cosmology
\\
\ \\
}}

Thomas Hartman, Yikun Jiang, and Edgar Shaghoulian

\vskip5mm
{\it Department of Physics, Cornell University, Ithaca, New York, USA
} 

\vskip5mm

{\tt hartman@cornell.edu, yj366@cornell.edu, eshaghoulian@cornell.edu }

\end{center}

\vspace{4mm}

\begin{abstract}
\noindent
A quantum extremal island suggests that a region of spacetime is encoded in the quantum state of another system, like the encoding of the black hole interior in  Hawking radiation. We study conditions for islands to appear in general spacetimes, with or without black holes. They must violate Bekenstein's area bound in a precise sense, and the boundary of an island must satisfy several other information-theoretic inequalities. These conditions combine to impose very strong restrictions, which we apply to cosmological models. We find several examples of islands in crunching universes. In particular, in the four-dimensional FRW cosmology with radiation and a negative cosmological constant, there is an island near the turning point when the geometry begins to recollapse. In a two-dimensional model of JT gravity in de Sitter spacetime, there are islands inside crunches that are encoded at future infinity or inside bubbles of Minkowski spacetime. Finally, we discuss simple tensor network toy models for islands in cosmology and black holes.

 \end{abstract}
%\vspace{.2in}
%\vspace{.3in}

\pagebreak
\pagestyle{plain}

\setcounter{tocdepth}{2}
{}
\vfill

\ \vspace{-2cm}
\renewcommand{\baselinestretch}{1}\small
\tableofcontents
\renewcommand{\baselinestretch}{1.15}\normalsize

\section{Introduction}

The holographic principle suggests that the entropy of a region in quantum gravity is bounded by its area in Planck units, 
\be\label{holobound}
S \leq \frac{\mbox{Area}}{4} \ .
\ee
For a static, spherically symmetric matter distribution, this follows from the Bekenstein energy bound $S \leq 2\pi RM$ and the threshold for black hole collapse, $M < \mbox{Area}/(8\pi R)$ \cite{Bekenstein:1980jp}. We will refer to \eqref{holobound} as the Bekenstein area bound.

For a QFT in flat spacetime, the Bekenstein energy bound follows from the positivity of relative entropy \cite{Casini:2008cr}. The status of the more general holographic bound \eqref{holobound} is not entirely clear. In dynamical, gravitating spacetimes, there are counterexamples for spacelike regions. Fischler and Susskind \cite{9806039} pointed out that it is violated by an arbitrary amount in FRW cosmology, because the area of a comoving region goes to zero near the big bang, while the matter entropy is constant. It is also violated in the interior of an evaporating black hole at late times, for a similar reason, with the large entropy provided by the interior partners of Hawking radiation. Even in Minkowski spacetime, it is violated by regions with null or nearly-null boundaries. These counterexamples led Bousso to conjecture a covariant bound on the classical matter entropy flux through a null surface \cite{hep-th/9905177, hep-th/0203101}. The Bousso bound evades the counterexamples and can be proved in special cases \cite{hep-th/9908070, hep-th/0303067, Bousso:2003kb, 1404.5635, 1406.4545}. 

Recent developments in the study of the black hole information paradox \cite{1905.08762, 1905.08255, 1908.10996, 1911.11977, 1911.12333} have led to a new interpretation of the black hole interior that depends, crucially, on the apparent violation of \eqref{holobound}. (See \cite{Almheiri:2020cfm} for a conceptual review.) The basic picture is that when \eqref{holobound} is violated near the black hole singularity, an `island' appears. The quantum state of the island, which covers most of the black hole interior, is secretly encoded in the Hawking radiation near null infinity. A sufficiently powerful observer collecting the radiation can in principle access the operator algebra in the interior. The boundary of the island is a quantum extremal surface (QES) \cite{1408.3203}, which is a surface of extremal generalized entropy. 
 The Bekenstein area bound \eqref{holobound} is violated by the semiclassical matter entropy in this situation, but it does not violate the spirit of the holographic principle -- the effective dimension of the Hilbert space associated to the interior region, as measured for example by our ability to entangle this region with an auxiliary system, is still set by the area. 

Since \eqref{holobound} is also violated in cosmology it is natural to ask whether there are quantum extremal islands.\footnote{In the real universe, a curious fact is that in our past lightcone, if we use the thermal entropy on the left-hand side, then the area bound \eqref{holobound} is violated around when the temperature reaches the TeV scale (despite no connection to electroweak physics!). }
 In this paper we explore several examples of crunching cosmologies where nontrivial islands indeed appear.  

The island suggests a relationship between the black hole interior and the Hawking radiation at null infinity similar to a holographic duality, though it does not necessarily entail a reduction in the spacetime dimension. It is holographic in the sense that it encodes the state of a gravitational system in a dual theory where gravity is unimportant. The encoding is similar to subregion duality in AdS/CFT, where the density matrix of a region on the boundary is encoded in the bulk entanglement wedge \cite{1211.3494, 1204.1330, 1408.6300}. This relationship is also known as entanglement wedge reconstruction. The examples in this paper support the idea that crunching regions in cosmology can be encoded holographically in non-crunching regions. 

In our first examples, the role of the Hawking radiation is played by a non-gravitating auxiliary system introduced to purify the thermal state of the matter fields in FRW. This suggests an interpretation of cosmological islands as a version of holographic duality where the island region is encoded in the quantum state of this auxiliary system. We also discuss  two-dimensional examples where instead of an auxiliary system, the quantum state in a subregion of dS$_2$ is encoded on ${
\cal I}^+$, or in a Minkowski bubble within dS$_2$. 

As noted in \cite{1908.10996}, the island proposal implies that an auxiliary qubit cannot be entangled with matter in a closed universe. If we tried to prepare such a state, the island would include the entire closed universe, and the entropy of the qubit would vanish despite our best attempt to entangle it. Our FRW examples replace the qubit by an entire QFT. In our setup the universe is infinite and the area term is non-zero, but it is overcome by the large matter entropy. Other cosmological applications of the island proposal have been considered recently in two dimensions in \cite{2007.02987,cgmTalks} and in higher-dimensional brane worlds in \cite{2007.06551}. See \cite{1711.01107,1901.04554,2002.11950} for other perspectives on holographic entanglement in de Sitter spacetime, and \cite{hep-th/0406134,hep-th/0503071,hep-th/0606204,1012.0274,1012.5302} for previous approaches to holography for crunching bubbles.

Before studying specific examples, we will study the general question of when quantum extremal islands, denoted by $I$, can exist in any given spacetime and quantum state. We discuss three simple necessary criteria: 
\begin{enumerate}
\item The Bekenstein area bound \eqref{holobound} must be violated by the island region, in a sense that we will make precise in section \ref{s:generalconditions}. The requirement is
\be
\Sred(I) \gtrsim \frac{1}{4} \Area(\p I) \ ,
\ee
where $\Sred$ is the finite part of the matter von Neumann entropy and the meaning of `$\gtrsim$' is discussed in detail below. The subtlety in deriving this formula is dealing with the UV divergences that would naively make the inequality trivial. 
 \item The boundary of an island must be in a \textit{quantum normal region} \cite{1905.08762}. This is by definition a region where the quantum expansion is positive in the outgoing direction and negative in the ingoing direction:
\be
\pm \frac{d}{d\lambda_{\pm}} S_{\rm gen}(I) \geq 0 \ ,
\ee
where $S_{\rm gen}$ is the generalized entropy. The derivatives are null deformations of the boundary of the island, with $d/d\lambda_+$ outgoing and $d/d\lambda_-$ ingoing with respect to $I$. 
\item Let $G$ be any region that surrounds the island, and shares a boundary with it (see figure \ref{fig:evap-regions} below for an example). Then 
\be
\pm \frac{d}{d\lambda_{\pm}} S_{\rm gen}(G) \leq 0 \ .
\ee
That is, the common boundary of $I$ and $G$ is also quantum normal with respect to $G$.
\end{enumerate}
A slightly different relationship between islands and the Bekenstein area bound was previously discussed in \cite{1908.10996}. Condition (2) is related to results of Engelhardt and Wall on quantum extremal surface barriers \cite{1408.3203} and it was derived for islands in \cite[\S 5]{1905.08762} as we will review below. One of our main observations is that the three conditions together are so strong that for practical purposes they are nearly sufficient to identify the islands in a given spacetime.

These criteria depend only on the island region and its immediate surroundings -- they make no reference to the choice of auxiliary region. To make this more explicit, let us choose a subsystem $R$ in a non-gravitating system and assume there exists an island $I$ in the gravitating region. In the black hole context, $R$ is the Hawking radiation, $I$ is (mostly) inside the black hole, and they are related by the large entanglement between interior and exterior Hawking pairs. The criteria above depend only on $I$, not on $R$. The statement is that if $I$ is the island associated to \textit{any} system $R$, then it obeys the three conditions.

In the black hole context, condition (2) implies that the boundary of the island must be outside (or on) the quantum apparent horizon. For eternal black holes, the quantum apparent horizon is the same as the event horizon, so it follows that the island ends outside (or on) the event horizon, as observed in \cite{Almheiri:2019yqk, Almheiri:2019psy}. For evaporating black holes, the quantum apparent horizon is inside the event horizon, and the island ends between them. In cosmology, the quantum normal region is typically inside the quantum apparent horizon. 

In every example we know of where all three conditions can be satisfied simultaneously, there are indeed nontrivial islands. In this sense the necessary conditions might actually be sufficient, too. This is just an empirical observation, with no derivation. It would be very interesting to derive sufficient conditions, especially for applications to higher dimensions, where the calculation of the matter entanglement entropy of disjoint regions, needed to find islands explicitly, is a serious technical challenge. In various big bang FRW cosmologies in four dimensions, we show by explicit construction of the islands that the conditions are sufficient.

In the rest of the introduction we will briefly summarize the various cosmological examples that we will study in the paper. In section \ref{s:review}, we review the island rule for entropy in gravitational systems. In section \ref{s:generalconditions}, we derive the general conditions (1), (2), and (3). We also discuss the requirement that islands must maximize the generalized entropy in all timelike directions, check the conditions in some previous examples of quantum extremal islands, and discuss the relation to the Bousso bound. In section \ref{s:matterentropy} we review properties of the matter entropy in FRW, and describe how to calculate the entropy of a region $I \cup R$ where $I$ is in FRW, and $R$ is in an auxiliary spacetime that purifies FRW. In sections \ref{s:radiation-only}-\ref{sec:JTglue}, we discuss the examples reviewed momentarily. In section \ref{s:tensors} we describe a tensor network toy model for islands, which also serves to highlight the similarities between black hole islands and cosmological islands. The tensor network model incorporates the fact that islands must violate the area bound but does not seem to capture the extremality condition or the quantum normal conditions in a natural way.

\subsection{Summary of examples}

In the examples we consider only spherically symmetric regions. The general conditions and most of our methods for FRW apply to regions of any shape.

\subsubsection*{FRW with radiation only}
We start with situations where the island $I$ is a region in a four-dimensional FRW cosmology. Region $R$, whose entropy we are calculating, is in an auxiliary Minkowski spacetime that purifies the matter in FRW. If the FRW universe is supported only by radiation, then there are no islands. This follows from the general conditions (1) and (2) --- the Bekenstein-violating region does not overlap with the quantum normal region in this cosmology. It also follows from condition (3) alone. Turning on a positive cosmological constant leaves these conclusions unchanged. See figures \ref{fig:radiation-only} and \ref{fig:positiveCC}.

\subsubsection*{FRW with radiation and negative CC}

\begin{figure}[t]
\begin{center}
\includegraphics{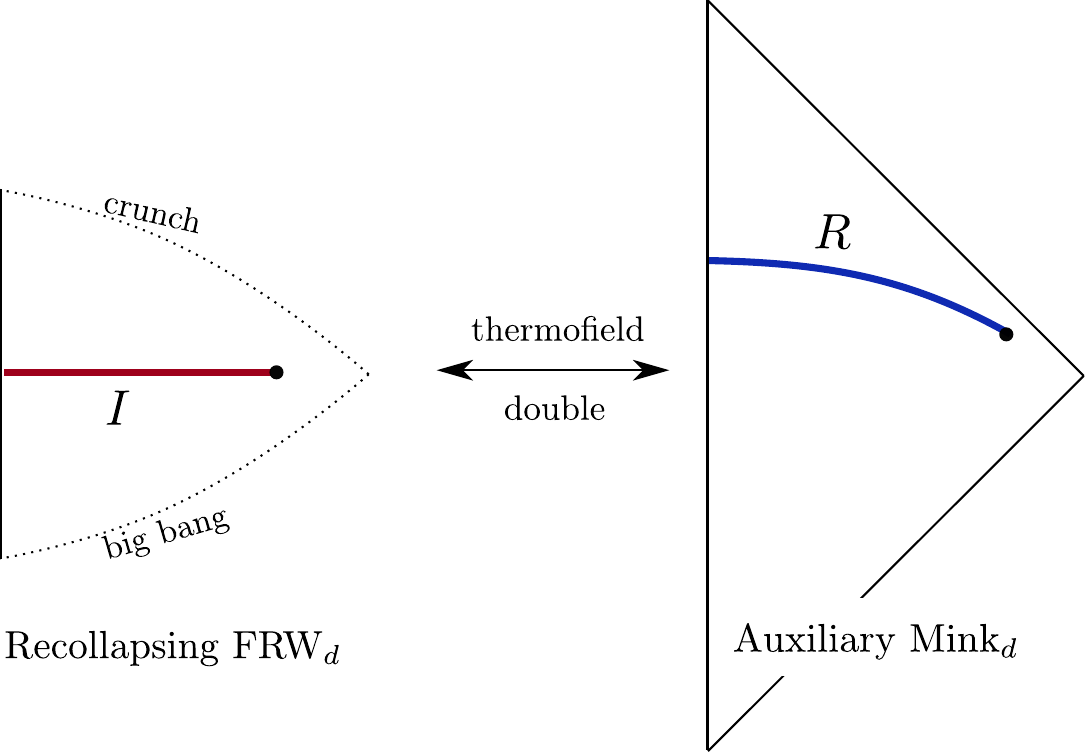}
\end{center}
\caption{A recollapsing FRW universe, with the thermal state of matter purified by an auxiliary Minkowski spacetime. We calculate the entropy of a large region $R$ in the Minkowski spacetime and find an island $I$ near the turning point of the FRW universe. \label{fig:recollapse-example}}
\end{figure}

If the FRW universe has a negative cosmological constant, the universe first expands, and then recollapses. In this case there are islands when region $R$ is large enough, as suggested by conditions (1)-(3) illustrated in figure \ref{fig:recollapse-regions}. The island $I$ always sits near the time of maximal scale factor, where the universe begins to recollapse. An example is illustrated in figure \ref{fig:recollapse-example}. If we set $t_R = 0$, and assume $r_R$ is large enough to violate the area bound, then the island is $I \approx R$. That is,
\be
t_I \approx t_R \approx 0 , \quad r_I \approx r_R \ .
\ee
If we increase $t_R$, the island stays at $t_I =0$, but it shrinks to have smaller radius.  The details of exactly how it shrinks depend on the matter sector; in a holographic CFT, we find an island with $r_I$  in the range
\be
r_R - \frac{v_B t_R}{a_0} \lesssim r_I \lesssim r_R
\ee
where $v_B$ is the butterfly velocity and $a_0$ is the maximal scale factor. 

This is one of the few cases where islands can be found analytically in higher than two dimensions. There is an interesting interplay with bounds on the matter entropy coming from the quantum null energy condition studied in \cite{1912.11024}.

\begin{figure}
\begin{center}
\includegraphics{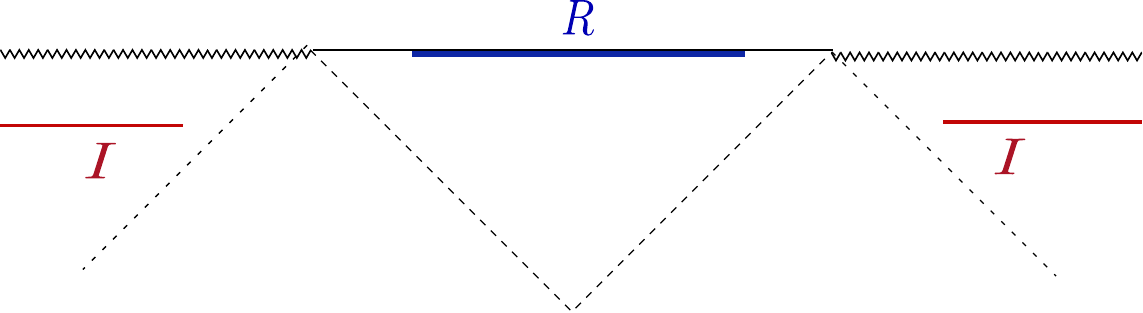}
\end{center}
\caption{Islands inside the crunching region of a 2d de Sitter model. \label{fig:intro-dS2}}
\end{figure}

\subsubsection*{JT gravity in dS$_2$}
Jackiw-Teitelboim (JT) gravity in dS$_2$ has a solution similar to the Schwarzschild-de Sitter black hole. We calculate the entropy of a region $R$ on ${\cal I}^+$, the spacelike future boundary, and find an island in the black hole region.  See figure \ref{fig:intro-dS2}. This is consistent with our conditions (1)-(3), which are illustrated in figure \ref{fig:dSregions}. This example is in fact very similar to the FRW thermofield double, because aside from the dilaton, the dS$_2$ black hole is two entangled copies of a hyperbolic FRW spacetime in the thermofield double state.

\begin{figure}
\begin{center}
\includegraphics{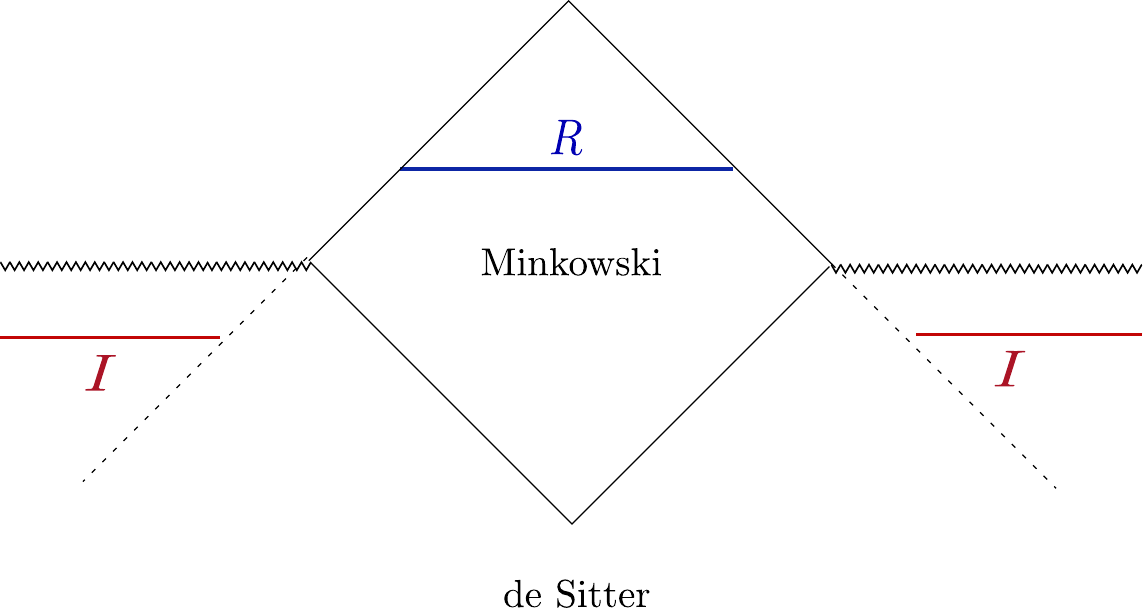}
\end{center}
\caption{Islands in 2d de Sitter spacetime with a Minkowski bubble and nearby crunching regions. \label{fig:intro-bubble}}
\end{figure}

\subsubsection*{Bubbles of flat spacetime in JT gravity in dS$_2$}
In the JT gravity calculation, the details at ${\cal I}^+$ do not play much role. This means we will get similar results if in this region we exit the de Sitter phase, for example by the nucleation of a flat-spacetime bubble. We model this in JT gravity by patching together de Sitter and Minkowski solutions, and calculate the entropy of regions inside the flat-spacetime bubble. Again we find islands in the ``crunching" region inside the black hole. See figure \ref{fig:intro-bubble}.  Our conditions (1)-(2) are completely independent of what happens outside of the hyperbolic patch where the island lives, as is condition (3) if we restrict $G$ to be in the same hyperbolic patch. In this case the constraints are the same as in the previous example, illustrated in figure \ref{fig:dSregions}. 

These two-dimensional examples have also been studied very recently in \cite{cgmTalks}, which overlaps with our sections 7 and 8.\footnote{In \cite{cgmTalks} the authors argued that bra-ket wormholes are necessary in this two-dimensional model to avoid paradoxes associated to islands timelike separated from region $R$. We will simply exclude timelike separated islands by hand, as it is not clear \textit{a priori} which saddlepoints should be included in the gravitational path integral (see also \cite{Anous:2020lka}). This produces an entropy with no obvious pathologies but it is possible that there are other contributions in some ranges of parameter space.
}

\section{Review of the island rule}\label{s:review}
Let $R$ be a non-gravitational system, such as a QFT, a subregion of a QFT, or a collection of qubits. Suppose we prepare $R$ in an entangled state with a gravitational system. The island formula \cite{1905.08762, 1905.08255, 1908.10996} computes the von Neumann entropy of system $R$, $S(\rho_R) = -\tr \rho_R \log \rho_R$. It states
\be\label{islandrule}
S(\rho_R) = \min \mbox{ext}_I \ S_{\rm gen}(I \cup R) \ ,
\ee 
where $I$ is a region in the gravitational theory --- the island --- and the generalized entropy is
\be
S_{\rm gen}(I \cup R) = \frac{\mbox{Area}(\p I)}{4} + \Ssemi(I \cup R)  - \Sdiv(\p I) \  .
\ee
$\Ssemi$ is the von Neumann entropy of the density matrix for the system $I \cup R$ as calculated in the semiclassical geometry (with fixed topology). $\Sdiv$ is the UV divergent part of the entropy associated to the boundary of region $I$. (It is often absorbed into the definition of the area term, but we will need to account for it explicitly.) The generalized entropy is extremized over the choice of $I$. If there are multiple extrema, including the trivial island $I = \varnothing$ which is always extremal, then we take the one with minimal $S_{\rm gen}$.

\begin{figure}[t]
\begin{center}
\includegraphics[scale=0.65]{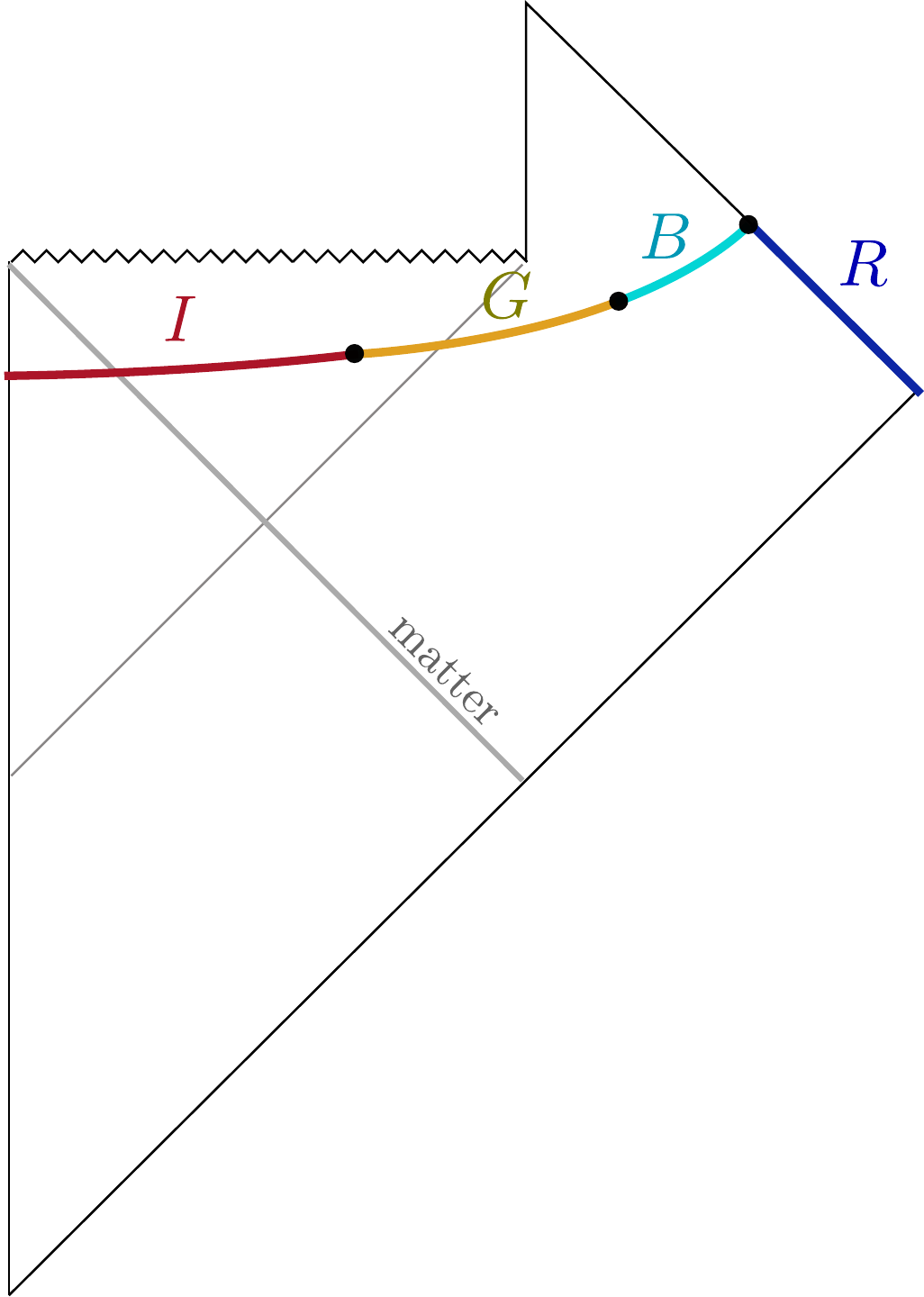}
\end{center}
\caption{\small Island $I$ in the interior of an evaporating black hole. \label{fig:evap-regions}}
\end{figure}

The formula \eqref{islandrule} for the entropy suggests that when there is a nontrivial island, the degrees of freedom in region $I$ are encoded in $R$. This can be formally derived to some extent using the technology of quantum error correction \cite{1411.7041, 1607.03901, 1601.05416}. To see why it makes sense, consider a Bell state with one qubit in region $R$ and its entangled partner in region $I$. This pair does not contribute to the entropy $\Ssemi(I \cup R)$, nor to $S(\rho_R)$, because the pair taken together is in a pure state. Since entanglement between $R$ and $I$ does not contribute to the entropy, we conclude that $I$ must be secretly encoded in $R$. In other words, operators in region $I$ can be rewritten as operators in $R$, though simple operators in $I$ will become very complicated and nonlocal under this map \cite{1911.11977, 1912.02210}.

The island rule is a generalized version of the Ryu-Takayanagi formula for holographic entanglement entropy \cite{hep-th/0603001, 0705.0016, 1304.4926, 1307.2892, 1408.3203, Barrella:2013wja}. It was discovered in an effort to understand the information paradox. As a black hole evaporates, the von Neumann entropy of the Hawking radiation increases. According to Hawking's calculation, it increases monotonically, and when the black hole evaporates completely, we are left with a finite entropy. This violates unitarity. The remarkable discovery of \cite{1905.08762, 1905.08255} is that at late times, when the paradox arises, there is a nontrivial island inside the black hole. This is illustrated in fig.~\ref{fig:evap-regions}. 
The island rule \eqref{islandrule} then gives a different formula for the radiation entropy, and this formula is compatible with unitarity.

Initially, the island rule was postulated on the basis that the radiation entropy should agree with the black hole entropy in a unitary theory \cite{1905.08762, 1905.08255}.
It was later derived by direct evaluation of the entropy by the replica method, using the path integral of semiclassical gravity \cite{1911.12333, 1911.11977}. It also has support from holographic arguments \cite{1908.10996, 1910.12836}. The path integral derivation, which builds on prior derivations of holographic entanglement entropy \cite{1304.4926,1705.08453}, requires a Euclidean (or Schwinger-Keldysh) path integral so it does not necessarily carry over to FRW cosmology. The replica calculations do apply to recollapsing FRW since there is a time reflection symmetry, but do not apply to all of the other examples in a straightforward way. In those cases we will just take the island proposal as a postulate.

\section{General conditions on islands}\label{s:generalconditions}

\subsection{Condition 1: The area bound is violated}\label{ss:condition1}
To form an island, we pay an entropy cost given by the island area in Planck units. This will only beat the trivial island if there is very high matter entanglement between $I$ and $R$. The trivial island $I = \varnothing$ leads to the entropy $S(R) = \Ssemi(R)$. Therefore for a nontrivial island to dominate, it must satisfy the extremality conditions and further obey
\be\label{islandinequality}
\frac{1}{4}\mbox{Area}(\p I)  + \Ssemi(I \cup R)  - \Sdiv(\p I) < \Ssemi(R) \ .
\ee
It follows that
\be\label{islandbek1}
\Ssemi(I) - \frac{1}{4}\mbox{Area}(\p I)  + \Sdiv(\p I) > \Ssemi(I) + \Ssemi(I \cup R) - \Ssemi(R) \ .
\ee
The right-hand side is positive by the Araki-Lieb inequality. If we temporarily ignore the divergences, this would seem to imply a violation of a Bekenstein-like bound, $\Ssemi(I) > \frac{1}{4}\Area(\p I)$. However this is trivial due to UV divergences -- the Araki-Lieb inequality applied to \eqref{islandbek1} does not constrain the finite part, it just requires $\Sdiv > 0$. 

Fortunately we can remove the divergences and obtain a nontrivial bound by a slightly more elaborate argument. First we will review the structure of divergences in the matter and gravitational entropies.

The generalized entropy is believed to be finite due to cancellations between the matter entropy and the geometric counterterms. See \cite[Appendix A]{1506.02669} for references and a review. To describe how this works, let us separate out the UV divergence associated to $\p I$ in the matter entropy by defining
\be\label{defsred}
 \Sred(I \cup R) = \Ssemi(I \cup R) - \Sdiv(\p I) \ ,
\ee
and similarly
\be
\Sred(I) = \Ssemi(I) - \Sdiv(\p I) \ .
\ee
In general, $\Sred(A)$ is defined by subtracting the UV divergences associated to components of the boundary $\p A$ in the gravitating region. We do not subtract divergences from the boundary of the non-gravitating region, $R$. The divergent piece $\Sdiv(\p I)$ is identical to the counterterm in the generalized entropy, so the generalized entropy of the island is finite,
\begin{align}
S_\textup{gen}(I) &= \frac{\Area(\p I)}{4} + \Ssemi(I) - \Sdiv(\p I) \\
&= \frac{\Area(\p I)}{4} + \Sred(I) \ . \notag
\end{align}
If the matter sector is a two-dimensional CFT, then
$\Sdiv = \frac{c}{6}N_p \log \frac{\epsilonrg}{\epsilonuv}$, where $N_p$ is the number of boundary points, $\epsilonuv$ is a UV length cutoff and $\epsilonrg \gg \epsilonuv $ is a renormalization length scale.\footnote{Most of the literature sets $\epsilonrg = 1$, or absorbs $\log \epsilonrg$ into $G_N$ and then sets $G_N=1$. We have kept it in order to see that logarithmic running won't affect the final result \eqref{matterbound2}.}  In $d>2$ spacetime dimensions,
\be
\Sdiv(\p I) \sim \# \frac{\mbox{Area}(\p I) }{ (\epsilonuv)^{d-2} } +  \cdots
\ee 
The dots are subleading divergences, including a logarithmic term $\log \frac{\epsilonrg}{\epsilonuv}$ in even dimensions. The coefficient of the leading term depends on the regulator, but its sign is fixed to be positive.

We now return to condition \eqref{islandinequality}, which can be restated in terms of the mutual information as
\be
\Isemi(I,R) \geq S_{\rm gen}(I)
\ee
with 
\begin{align}
\Isemi(I,R) &= \Ssemi(I) + \Ssemi(R) - \Ssemi(I \cup R) \\
&= \Sred(I) + \Sred(R) - \Sred(I \cup R) \ . \notag
\end{align}
For any region $R'$ containing $R$, strong subadditivity requires
\be
\Isemi(I, R') \geq \Isemi(I,R) \ , 
\ee
and so
\be\label{irss}
\Isemi(I,R') \geq S_{\rm gen}(I) \ .
\ee
\begin{figure}
\begin{center}
\includegraphics{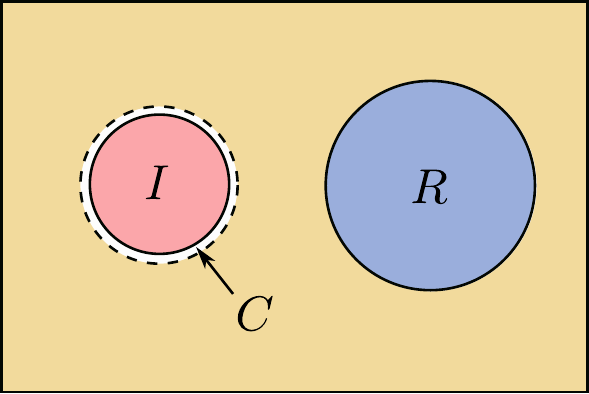}
\end{center}
\caption{Regions used to regulate the Bekenstein area bound. $I$ is the island, $R$ is the non-gravitational system appearing in the island formula, and $C$ is the narrow region of width $\delta$.\label{fig:ssa-regions-2}}
\end{figure}
To turn this into a constraint that refers only to region $I$, we define a narrow region $C$ that surrounds $I$, and pick $R' = (I \cup C)^\setcomp$. This is illustrated in figure \ref{fig:ssa-regions-2}. We may assume the full state is pure by including a purifying system in $R'$. With these choices the quantity $\Isemi(I,R')$ is the mutual information regulator for entanglement entropy introduced in \cite{1506.06195}. Assuming the width $\delta$ of region $C$ to be small (but much larger than $\epsilonuv)$, we have
\begin{align}
\Isemi(I, (I \cup C)^\setcomp) &= \Sred(I) + \Sred(I \cup C) - \Sred(C) \\
&\approx 2\Sred(I)  - \Sred(C) \ ,\notag
\end{align}
up to corrections that vanish as $\delta \to 0$.
Therefore \eqref{irss} becomes
\be\label{irss2}
\Sred(I) \gtrsim \frac{1}{4}\Area(\p I) + \Sred(C) \ .
\ee
The entropy of the narrow region $C$ takes the form
\cite{0905.2562,1202.2068}
\be
\Ssemi(C) = \Sdiv(\p I) + \Sdiv(\p I^+) - \kappa \frac{\Area(\p I)}{\delta^{d-2}}  + \cdots , 
\ee
where the dots are subleading and $\kappa$ is a scheme-independent constant that depends on the matter content. In two dimensions, $\delta^{d-2}$ is replaced by a log. Although the last term looks similar to a UV divergence, this term is physical (i.e. universal), because $\delta$ is a physical length scale in the setup, not the UV cutoff.  The finite part is
\be
\Sred(C) \approx - \kappa \frac{\Area(\p I)}{\delta^{d-2}}  \ .
\ee
The area term appearing here is much smaller than the area measured in Planck units, since $\delta \gg \ell_P$, so it can be neglected in \eqref{irss2}.  Therefore we have derived the necessary condition
\be\label{matterbound2}
\setlength\fboxsep{0.25cm}
\setlength\fboxrule{0.4pt}
 \boxed{ 
\Sred(I) \gtrsim \frac{\Area(\p I)}{4} \ .
}
\ee
The matter entropy is the finite part of the von Neumann entropy. The notation `$\gtrsim$' means that we should only take seriously terms that are the same order as (or larger than) the right-hand side, because of the approximations made in the derivation.

The conclusion is that the finite part of the quantum entropy of region $I$ must violate the Bekenstein area bound.  As we have emphasized in the introduction, this condition refers only to the island region $I$, so it must be satisfied by the island for any choice of $R$. 

A closely related condition can be stated that references a region $G$ surrounding $I$, in the spirit of condition (3) in section \ref{sec:qnormal2}. We simply rewrite \eqref{irss} with $G = (I \cup R')^c$ as  $S_{\rm gen}(I \cup G) \geq S_{\rm gen}(G)$. This constrains the generalized entropy under growing region $I$ by appending region $G$. For $R' = R$ this is equivalent to the dominance condition, although we will sometimes constrain region $G$ to be smaller, which will give a weaker condition. Taking $G \rightarrow 0$ and expanding in terms of the matter entropy is what gave \eqref{matterbound2} above.

If the matter in region $I$ is in a thermal state, then the extensive part of $\Sred(I)$ is equal to the thermodynamic entropy of region $I$. In other states $\Sred(I)$ can be much smaller than the thermodynamic entropy. An example of this effect is an excited state produced by a global quantum quench at $t=0$, with region $I$ taken to have size $L \gg t$ \cite{cond-mat/0601225, 0708.3750, 1303.1080, 1305.7244, 1311.1200}. This is also relevant to cosmology because the quantum state produced by reheating at the end of inflation is like that of a global quench -- if the inflaton is in a pure state for $t<0$ and inflation ends at $t=0$, then the matter supporting the FRW solution for $t>0$ is thermal on subhorizon scales but purified on longer distances (see e.g. \cite{1003.3011} for a related discussion).

\subsection{Condition 2: $ I$ is quantum normal}\label{sec:qnormal}

If the island has a boundary, then it also obeys the extremality condition
\be\label{econ}
\frac{d}{d \lambda}  S_\textup{gen}(R \cup I)  = 0 \ .
\ee
We will take the derivative in a null direction. That is, let $X^\mu(\sigma)$ be the embedding functions defining the surface $\p I$ and $k^{\mu}(\sigma)$ be a null vector field normal to $\p I$, specifying the profile of the deformation along the surface.
We define the derivative $\frac{d}{d\lambda}$ by deforming 
\be
X^\mu(\sigma) \to X^\mu(\sigma) + \lambda k^\mu(\sigma)\,.
\ee
At various points we will use the notation $d/d\l_+$ to refer to outward null derivatives and $d/d\l_-$ for inward null derivatives with respect to region $I$.

By adding and subtracting $\Ssemi(I)$ to \eqref{econ} we find 
\be\label{ddl}
\frac{d}{d\lambda}\left[ \Ssemi(I) + \frac{\Area(\p I)}{4}-\Sdiv(\p I) \right] + \frac{d}{d\lambda}\left[ \Ssemi(I \cup R) -\Ssemi(I) - \Ssemi(R) \right] = 0 \ .
\ee
We have also added the term  $\frac{d}{d\lambda}\Ssemi(R)$, which vanishes because the deformation does not affect region $R$.  Each term in brackets is UV-finite. 

The first term in brackets is the generalized entropy of the island, and the second is the mutual information, up to a sign.
Thus we can rewrite the extremality condition as
\be\label{extrI}
\frac{d}{d\lambda} S_{\rm gen}(I) = \frac{d}{d\lambda} \Isemi(I,R) \ .
\ee
This version of the extremality condition has the advantage that both $S_\textup{gen}$ and the mutual information are UV-finite. It is also in this form that conditions (2) and (3) have a simple physical explanation. The mutual information measures correlations between $I$ and $R$. Intuitively, it should be impossible for the mutual information to change too rapidly, because there is only a finite amount of matter near $\p I$ that can potentially be correlated with a region elsewhere. Therefore we expect both upper and lower bounds on $\frac{d}{d\lambda}S_{\rm gen}(I)$ that depend only on what matter is present near $\p I$. 

The lower bound was obtained in \cite{1905.08762} as follows. Let us choose the direction of increasing $\lambda$ to be an outward null direction, so we denote $\lambda = \lambda_+$.  That is, the original region is a subregion of the deformed region. With this convention, strong subadditivity of the matter entropy is equivalent to monotonicity of the mutual information with the sign
\be
\frac{d}{d\lambda_+} \Isemi(I,R) \geq 0 \ .
\ee
For inward null deformations the inequality is reversed. Therefore, for (past or future directed) null deformations,
\be\label{derivbound}
\setlength\fboxsep{0.25cm}
\setlength\fboxrule{0.4pt}
\boxed{
\pm\frac{d}{d\lambda_{\pm}} S_{\rm gen}(I) \geq 0 \ .
}
\ee
 This derivative of the generalized entropy is proportional to the quantum expansion \cite{1506.02669}, so we can also write the condition as
\be
\Theta_{\pm +} \geq 0 , \quad \Theta_{\pm -} \leq 0 \ .
\ee
Here we are using the notation of \cite{hep-th/0203101}, where the first/second sign denotes inward $(-)$ or outward $(+)$ in the time/space direction. We say that a region $ I$ obeying these conditions lives in a \textit{quantum normal region}, in analogy with the ordinary normal region defined by the classical expansion (see \cite{hep-th/9905177, hep-th/9906022}).

Note that `outward' is defined with respect to the island. If the island has multiple boundaries, then the notion of outward depends on the boundary.

\subsection{Condition 3: $G$ is quantum normal}\label{sec:qnormal2}
The physical intuition below equation \eqref{extrI} suggests that there is also an upper bound on $\frac{d}{d\lambda_+}S_{\rm gen}(I)$. This is our condition (3).
Let $G$ be a region that surrounds the island and shares a boundary, as in figure \ref{fig:evap-regions}. $G$ can be infinite or it can end at another boundary, but we assume it is spacelike separated from $R$ (or more accurately, achronal with respect to $R$). The third general condition is 
\be\label{condition3}
\setlength\fboxsep{0.25cm}
\setlength\fboxrule{0.4pt}
\boxed{
\pm \frac{d}{d\lambda_{\pm} }S_{\rm gen}(I) \leq \pm \frac{d}{d\lambda_{\pm} }\left[ \Ssemi(I) - \Ssemi(G) \right]\,.
}
\ee
Since $I$ and $G$ share a boundary, and it is only along this shared boundary that the deformation affects that area term, this can also be written
\be
\pm \frac{d}{d\lambda_{\pm} }S_{\rm gen}(G) \leq 0 \  .
\ee
The deformation $d/d\lambda_+$ is outgoing with respect to $I$, but ingoing with respect to $G$, so this says that $G$ must also be quantum normal.

This third condition is on a slightly different footing from the first two because it involves the choice of region $G$ outside the putative island. The condition becomes stronger for larger $G$, but we cannot take $G$ too large because we assumed $G$ is spacelike separated from $R$. Once we have picked a region $G$, the statement is that condition (3) applies to any island coming from a region $R$ achronal with respect to $G$. This will be particularly useful in our cosmological examples where we have two completely separate spacetimes, one with gravity and one without. Region $G$ will be chosen as the portion of the gravitating spacetime that is not in $I$.

To derive condition (3) we start with the extremality condition in the form 
\be\label{mmex}
\frac{d}{d\lambda_\pm} S_{\rm gen}(I) = \frac{d}{d\lambda_\pm}\left[ \Ssemi(I) - \Ssemi(I \cup R) \right] \ .
\ee
For any region $R'$ containing $R$, strong subadditivity of the matter entropy implies
\be
\pm \frac{d}{d\lambda_\pm}\Ssemi(I \cup R') \leq \pm \frac{d}{d\lambda_\pm}\Ssemi(I \cup R) \ .
\ee
Replacing $I \cup R'$ by its complement (including a purifying system if necessary) and plugging into \eqref{mmex} gives the third condition, \eqref{condition3}.

\subsection{Comments on the three conditions}\label{sec:comments}

As emphasized in the introduction, the three conditions refer only to region $I$ and its vicinity. They must be satisfied by the island associated to any region $R$ that is spacelike separated from $I$ and $G$.

 Together conditions (2) and (3) can be restated as 
\be\label{fattening}
\mp\frac{d}{d\lambda_\pm} \Sred(I) \leq \pm \frac{1}{4} \frac{d}{d\lambda_\pm} \Area(\p I) 
\leq  \mp\frac{d}{d\lambda_\pm} \Sred(G) \ .
\ee
In the classical limit, the upper and lower bounds coincide, so this reduces to the usual extremality condition of an HRT surface \cite{0705.0016} $\frac{d}{d\lambda_{\pm}}\Area(\p I) = 0$. In the quantum case, the allowed region fattens out as prescribed by \eqref{fattening} to become codimension-0.

As we derived them, the conditions above only apply when $I$ is disjoint from $R$. An example where $I$ and $R$ are connected occurs for an eternal black hole in AdS with left and right boundaries coupled to flat space baths. In this case, we can have region $R = R_{\rm left} \cup R_{\rm right}$, where $R_{\rm left}$ consists of the entire left flat space bath including the left AdS boundary, and $R_{\rm right}$ consists of a portion of the right flat space bath that does not include the right AdS boundary. The quantum extremal surface simply extends $R_{\rm left}$, i.e. $R_{\rm left} \cup I$ can be continuously deformed into $R _{\rm left}$, as shown in figure \ref{fig:eternalBHads}. In this case, our condition (1) does not apply, condition (3) applies unchanged, and condition (2) can be shown to apply to region $R_{\rm left} \cup I$ using the same argument as before.

It is natural to ask whether the conditions are also sufficient for islands. We do not know the answer but we are not aware of any counterexamples. That is, in spacetimes with a finite region satisfying all three conditions, there seems to be an island somewhere in this region.  In fact, this is true even if we drop condition (3). 

There is a sense in which the Bekenstein condition is in tension with the other two, so for them to all be satisfied requires special circumstances. To explain this, note that we can always make the area of the boundary of a region $I$ arbitrarily small while hardly affecting the region itself, by wiggling the boundary in the time direction to make $\p I$ nearly null. This is one reason why Bekenstein's area bound needed to be upgraded to the Bousso bound to formulate a reasonable covariant entropy bound \cite{hep-th/9905177, hep-th/0203101}.

Choosing a wiggly boundary makes it easy to violate the Bekenstein area bound, but comes at the expense of making it more difficult to find a quantum normal region. This is because introducing wiggles in a timelike direction increases the classical expansion at the `bottom' of a wiggle, and decreases the classical expansion at the `top' of a wiggle. For example, in flat spacetime, the wiggly region bounded by the surface with $\{ t=t(x_1), x_2=x_3 = \cdots = 0\}$, has null extrinsic curvature proportional to $t''(x_1)$. If the wiggles are very sharp, then in any spacetime this will make the sign of $\frac{d}{d\lambda} \Area$ oscillate as we go around the boundary of region $I$, making it difficult to satisfy the quantum normal conditions.

There is also tension between conditions 2 and 3, because as described above they can only be satisfied classically when the inequalities are saturated. Thus taken together the three conditions are very restrictive, so if they are all satisfied, this is a strong hint that an island can be found nearby.

\subsection{The island is a maximum in the time direction}\label{ss:maximin}

The extremality condition does not explicitly refer to whether the quantum extremal surface is a local maximum, minimum, or saddlepoint of the generalized entropy. However it is believed that the QES must always be a saddlepoint, and this can be proved under some additional assumptions. This can be useful to search for islands.

In the classical case, Wall gave an alternative definition of the extremal surface as the maximin surface obtained by minimizing the area on a slice $\Sigma$, then maximizing over the choice of $\Sigma$ \cite{1211.3494}. Together with focusing, this implies that the area of an extremal surface is always maximal in any timelike direction. That is, the area is non-increasing at second order under timelike deformations. To simplify the notation we will explain this in two dimensions. Let us expand the area term (i.e. the dilaton) to second order around the extremal surface,
\be
\Area  = \phi(0) + \alpha (x^+)^2 + \beta(x^-)^2 + \gamma x^+x^- \ .
\ee
Here $x^\pm = t \pm x$ are local null coordinates. The classical focusing equation $\nabla_{\pm}^2 \phi \leq 0$ implies $\alpha, \beta \leq 0$. The maximin prescription implies that there exists a spacelike direction, $\p_x + a \p_t$ with $|a|<1$, in which the area is increasing. Together, these require $\gamma\leq 0$. We can then choose any timelike direction and use the relations $\alpha,\beta,\gamma \leq 0$ to conclude that the area is non-increasing along this direction at second order.

In the quantum case, the maximin prescription still applies \cite{1912.02799} if we also assume the quantum focusing conjecture \cite{1506.02669}. Thus quantum extremal surfaces are also maximal under all timelike deformations.  We give a direct derivation of this statement in appendix \ref{app:maximin} that uses quantum focusing but does not rely on maximin. The idea of the derivation is to take another derivative of the extremality condition \eqref{extrI} with respect to the boundary of region $R$. We then combine strong subadditivity of the matter entropy, quantum focusing, and entanglement wedge nesting to show $\gamma \leq 0$ in the quantum case. 

We will study an example in section \ref{ss:qnecbound} where requiring the island to be  a maximum under timelike deformations places nontrivial constraints on the matter entropy. Interestingly, the resulting constraint turns out to be the same as the constraint derived from the quantum null energy condition (QNEC) in \cite{1909.00919}.

\subsection{Examples}
\label{ss:examples}
In this subsection we will check the necessary conditions from above in a few examples. These examples are of islands that have previously appeared in the literature, so it is simply a check of our conditions above before we apply them to novel cases. 

The first example is when the state of a closed universe is pure up to some amount of matter entangled with matter in a separate universe. In this case the island region always includes the entire closed universe. The violation of the Bekenstein area bound $\Sred(I) \gtrsim$ Area$(\p I)/4$ occurs because the island has no boundary, implying Area$(\p I) = 0$, while there is some nonvanishing matter entanglement. The requirement that the quantum extremal surface be quantum normal with respect to regions $I$ and $G$ is trivial, since it is a statement about the endpoints of the island, which do not exist in this case. 

\begin{figure}
\begin{center}
\includegraphics[scale=0.35]{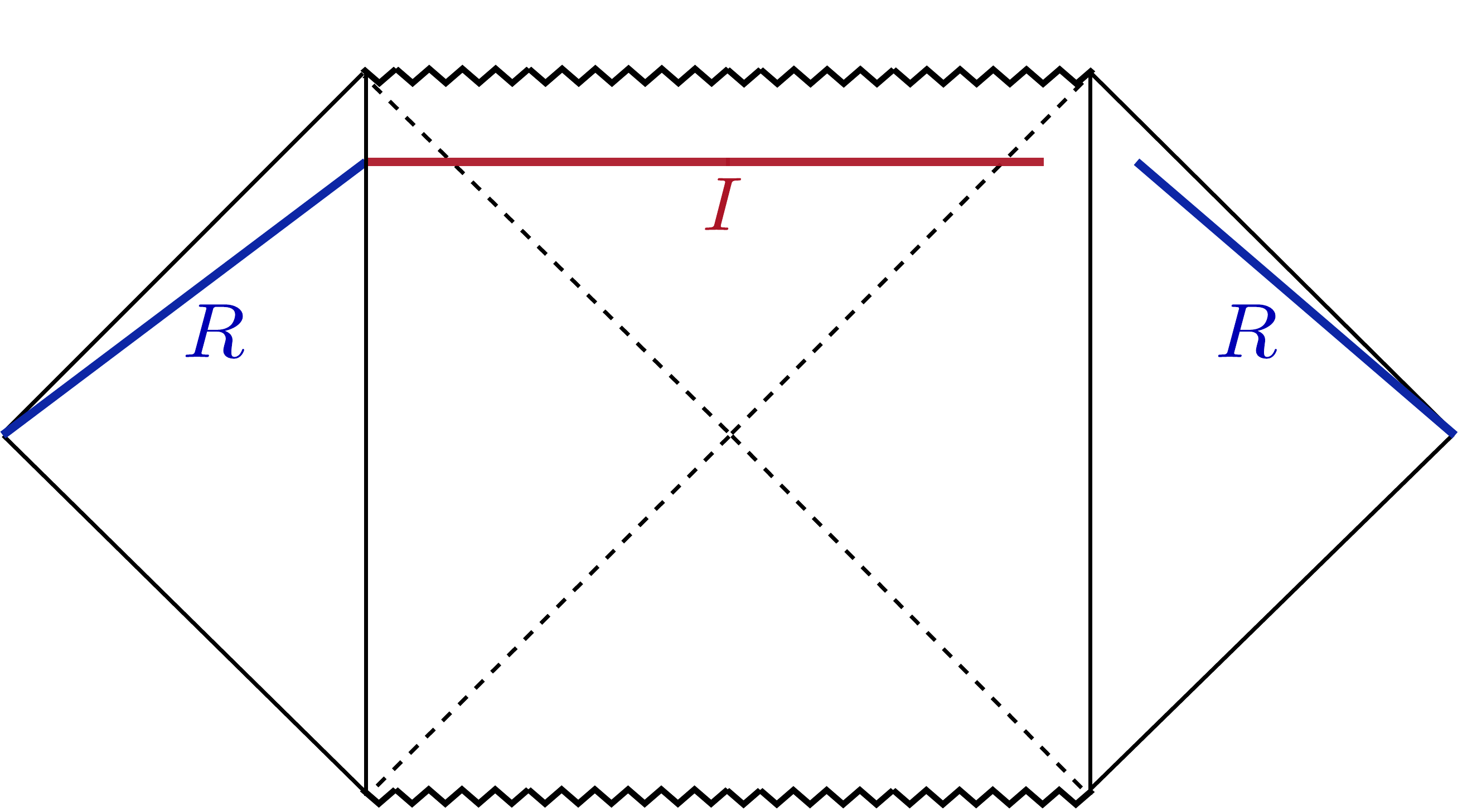}
\end{center}
\caption{$I$ is an ``island" (sometimes called a peninsula because it is connected to $R$) inside an eternal black hole for region $R$.\label{fig:eternalBHads}}
\end{figure}

The next example is for the eternal black hole in AdS spacetime, glued to asymptotic flat regions as in figure \ref{fig:eternalBHads}. In this case, taking the entire left flat spacetime and part of the right flat spacetime as region $R$, the quantum extremal surface is outside of the right horizon. This is a special case where there is not a disconnected island region, as discussed in section \ref{sec:comments}, so our condition (2) has to be modified. To compute the quantum apparent horizon, we pick a region that begins at spatial infinity on the left and ends at some point outside the right horizon. Due to the time translation symmetry, the quantum apparent horizon for this region coincides with the event horizon. This conclusion is independent of dimensionality and asymptotics, since it follows directly from the timelike Killing vector. This example explains why the island is outside the event horizon for eternal black holes \cite{Almheiri:2019yqk}. The argument also applies to the finite-size islands in eternal black holes studied in AdS in \cite{Almheiri:2019yqk, Almheiri:2019psy} and Minkowski spacetime in \cite{2020arXiv200400598F, Anegawa:2020ezn, Hartman:2020swn}, because they are in a regime where the problem factorizes into two copies of the situation just discussed.

\begin{figure}
\begin{center}
\includegraphics[scale=0.35]{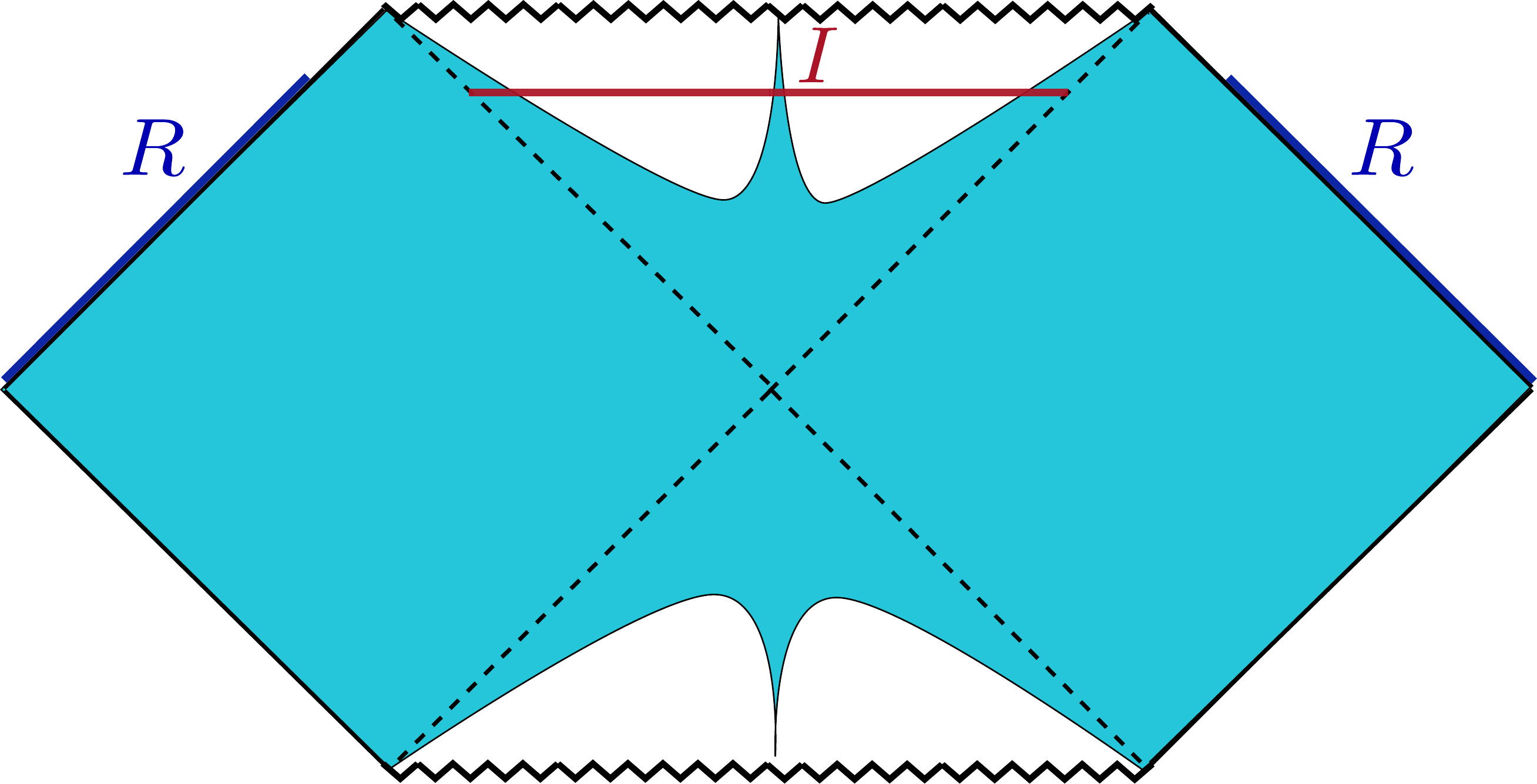}
\end{center}
\caption{$I$ is an island inside an eternal black hole for region $R$. The quantum normal region for $I$ is shaded.\label{fig:eternalBH}}
\end{figure}

It is straightforward to check the Bekenstein area bound and the quantum normal conditions in the context of the CGHS/RST model in two-dimensional flat spacetime. Islands in this model were found in \cite{2020arXiv200400598F, Anegawa:2020ezn, Hartman:2020swn}. For an eternal black hole, the position of the island in relation to the quantum normal region for $I$ is depicted in figure \ref{fig:eternalBH}. It is outside as required. Details for this case as well as the evaporating black hole in the CGHS/RST model can be found in appendix \ref{app:cghs}.

Condition (3) in this context is more subtle. We must first choose a region $G$ that extends outside the island, to some point $x_G$. Then condition (3) applies to any island associated to a region $R$ that is spacelike separated from $G$. If the point $x_G$ is outside the event horizon then this places a cutoff on the allowed region $R$ along ${\cal I}^+$. A natural location for $x_G$ that does not restrict the region $R$ is near the evaporation endpoint. As shown in appendix \ref{app:cghs}, the constraint from this choice is severe enough that the actual QES sits on the boundary of the allowed region.

\subsection{Comments on the Bousso bound}

Certain entropies associated to a null cone are constrained by the classical and quantum Bousso bounds \cite{hep-th/9905177, hep-th/0303067, hep-th/9907062, 1506.02669}. We would like to see how this is compatible with the violation of the Bekenstein area bound for region $I$. 

Conceptually there are two different ways that we can have $\Ssemi(I) > \frac{1}{4}\Area(\p I)$ while satisfying the Bousso bounds. The first is if region $I$ cannot be deformed to a null surface --  for example, a large region at early time in FRW violates the Bekenstein bound but not the Bousso bound (which is the Fischler-Susskind bound in this context \cite{9806039}), because the singularity prevents us from deforming $I$ to a null cone. The second mechanism is that when the entropy is intrinsically quantum, the Bousso bound gives only a lower bound, not an upper bound. This is the mechanism at play inside an old black hole. In this case, the von Neumann entropy of region $I$ is equal to the von Neumann of its past lightcone, because they have the same causal domain, but the entropy is quantum mechanical so it has no upper bound.

\begin{figure}[t]
\begin{center}
\includegraphics{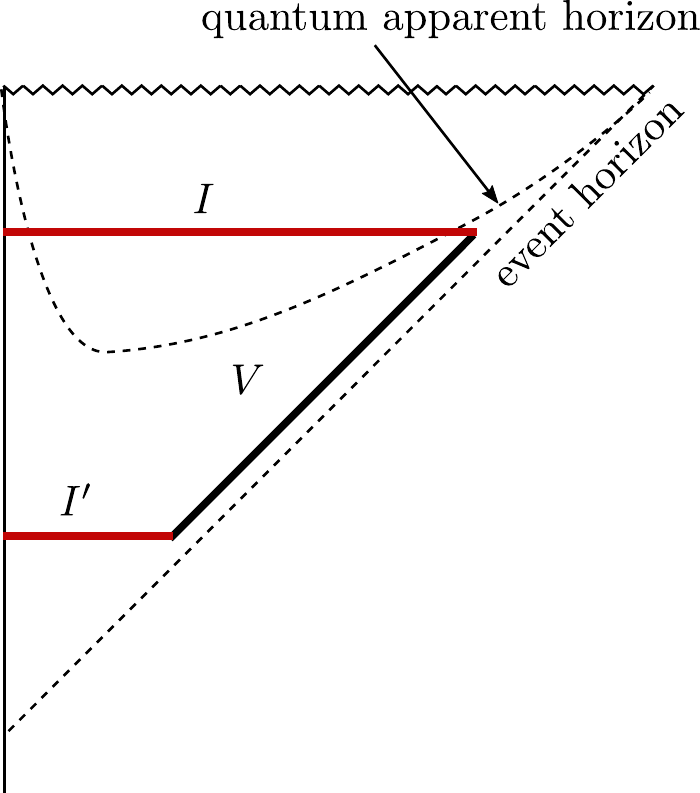}
\end{center}
\caption{$I$ is an island inside an evaporating black hole. The Bousso bound applied to $I,I'$ and $I^\setcomp, (I')^\setcomp$ places an upper bound on the flux of classical entropy through $V$ and the quantum entropy of $I'$, but it does not put an upper bound on the quantum entropy of $I$.\label{fig:boussoregions}}
\end{figure}

Referring to figure \ref{fig:boussoregions}, the classical Bousso bound states that the flux of hydrodynamic entropy through the null surface $V$  is bounded above by $\frac{1}{4}\Area(\p I) -\f 1 4 \Area(\p I')$. The hydrodynamic entropy is classical entropy (from, say, a cup of tea) that can be calculated by the integral of a local density; it contributes to the von Neumann entropy of $V$, but it does not include the quantum effects that produce most of the entropy inside an old black hole. So this is not a contradiction.

The quantum Bousso bound does constrain the full von Neumann entropy, but it is also compatible with $\Ssemi(I) > \frac{1}{4}\Area(\p I)$. The quantum bound requires \cite{1506.02669}
\be\label{qbousso}
\Ssemi(I') - \Ssemi(I) \leq \frac{1}{4}\Area(\p I) - \frac{1}{4}\Area(\p I') 
\ee
and
\be\label{qbousso2}
\Ssemi((I')^\setcomp) - \Ssemi(I^{\setcomp}) \leq \frac{1}{4}\Area(\p I) - \frac{1}{4}\Area(\p I') \ .
\ee
Neither of these inequalities places an upper bound on the quantum entropy  $\Ssemi(I)$. The latter inequality is responsible for the classical bound as a special case when a cup of tea falls through $V$.

This discussion has a simple consequence for how islands can be created. Let us consider an experiment where we try to engineer a faraway island by manufacturing entangled qubits in a lab, then sending half of the qubits to a distant region of space. Can the qubits be carefully arranged in a way that produces an island, allowing us to access this region of space from the safety of our lab? The answer is no if we assume the Bousso bound holds. In performing this experiment, the qubits would cross a null surface where we can invoke the classical Bousso bound. (Since the entangled qubits are widely separated from their partners, they will contribute to the hydrodynamic entropy flux that appears in the classical Bousso bound, despite the quantum origin of their entropy.) It is therefore impossible to create enough entanglement to violate the Bekenstein area bound in this manner. The entanglement responsible for an island must be created in a more subtle way, as it is in Hawking radiation. On the other hand, perhaps transporting entangled qubits can be used to enhance the island effect, producing islands in situations with other sources of entanglement but where they would not appear naturally.

\section{Matter entropy in FRW}\label{s:matterentropy}

\subsection{Thermofield double setup}

Consider a spatially flat FRW cosmology in $d$ spacetime dimensions,
\be\label{frwcoords}
ds^2 = -dt^2 + a(t)^2 dx^2 \ .
\ee
Conformal coordinates are defined by
\be\label{conformalcoords}
ds^2 = a(\eta)^2( -d\eta^2 + dx^2) , \quad \eta(t) = \int_{0}^{t} \frac{dt'}{a(t')} \ .
\ee
We assume the matter is in a thermal state. Although this is standard in cosmology, the assumption we are making here is actually much stronger than the usual one, because standard cosmology is not sensitive to the microscopic details of the quantum state of the matter --- it applies just as well to a pure state that is approximately thermal on distance scales larger than the thermal correlation length. Our setup by contrast assumes the microscopic quantum state of the matter to be mixed on much larger scales.

Like any mixed state, the thermal state of the radiation can be purified by doubling the degrees of freedom. A convenient (though highly non-unique) way to do this is the thermofield double. Introduce a second copy of the matter QFT on an auxiliary, non-dynamical $d$-dimensional Minkowski spacetime, where the metric is
\be
ds^2 = -d\eta^2 + dx^2 \ .
\ee
We use the same coordinate labels $(\eta, x)$ but we will always specify whether we are referring to the original spacetime or the auxiliary, purifying spacetime.
The thermofield double is the pure state 
\be\label{tfd}
|\beta_0\rangle =\frac{1}{\sqrt{Z}} \sum_{n} e^{-\beta_0 E_n/2}|n\rangle_1^* |n\rangle_2 \ ,
\ee
where $*$ denotes CPT conjugation. Upon tracing out one copy, the reduced density matrix in the other copy is thermal.

There is no gravity on the auxiliary space. This means that we are not purifying the graviton radiation in the original FRW. We can ignore this issue by assuming the graviton entropy is subleading compared to other components (which it is in the real world).

\begin{figure}
\begin{center}
\includegraphics[scale=1]{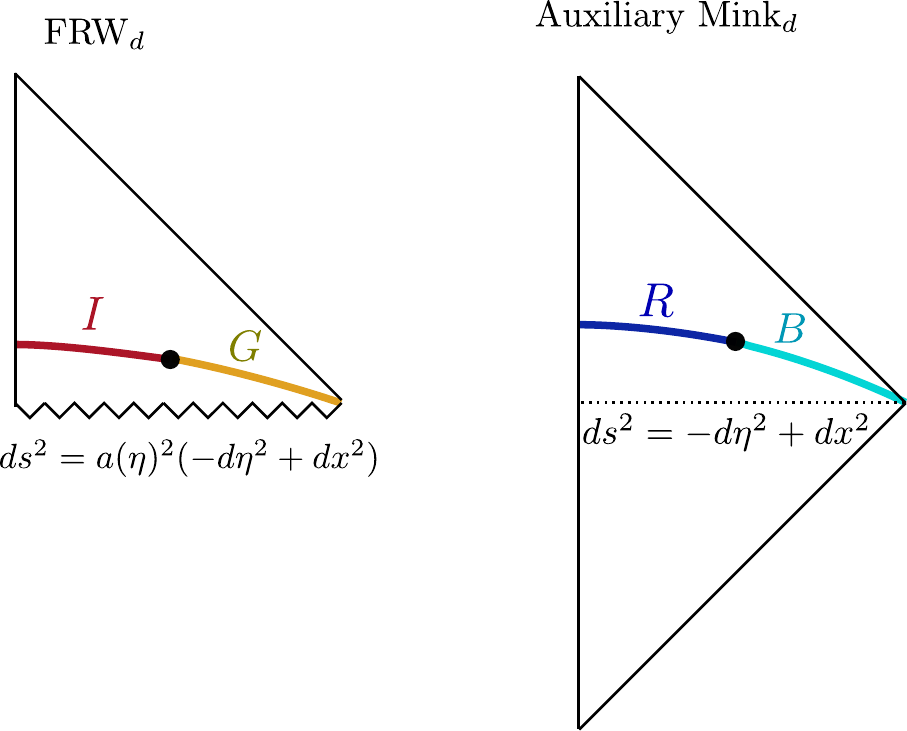}
\end{center}
\caption{ Regions in FRW and the auxiliary purifying spacetime. The regions $R,B,G,I$ will play roles similar to the regions with the same labels in the evaporating black of figure \ref{fig:evap-regions}. \label{fig:frwisland}}
\end{figure}

To define the thermofield double state that purifies FRW, we first prepare the thermofield double in two copies of Minkowski spacetime,
\be\label{tfdmm}
ds_1^2 = -d\eta^2 + dx^2 \quad \mbox{and} \quad ds_2^2 = -d\eta^2 + dx^2 \ .
\ee
This is prepared by a Euclidean path integral (for the matter fields only) on a strip of length $\beta_0/2$.
Then we do a conformal transformation that inserts the conformal factor $a(\eta)^2$ in copy 1 but acts trivially in copy 2. The inverse temperature in the thermofield double is denoted $\beta_0$ to distinguish it from the physical inverse temperature $\beta = a \beta_0$ in FRW.

We will study the entropy of a region $R$ in the auxiliary spacetime. This region is specified by a spatial region $\Sigma_R$ at fixed time $\eta_R$. $R$ plays the same role that the Hawking radiation played in the evaporating black hole. 

We will look for an island in the gravitating FRW region. First we need to understand properties of the matter entanglement when region $I$ is in the FRW spacetime and region $R$ is in the auxiliary Minkowski spacetime. We will do this in two steps --- first we will consider the matter entanglement in two copies of Minkowski spacetime, then we will turn on the scale factor. We assume the matter sector is described by a CFT and that the number of degrees of freedom in the CFT is large enough that we can ignore the entropy of gravitons.

A natural way to realize the FRW thermofield double is through a Randall-Sundrum brane construction in higher-dimensional AdS. This may provide another way to analyze this model along the lines of \cite{2019JHEP...07..065C, 1908.10996, 2020arXiv200604851Z, Balasubramanian:2020hfs, Chen:2020jvn}, but we will not take this perspective.

\subsection{Two copies of Minkowski}
We start by analyzing the matter entanglement entropy for a thermofield double state of two copies of flat spacetime. To distinguish this from FRW we use a tilde,
\begin{align}
\Sflat &\equiv \mbox{von Neumann entropy in two entangled copies of Minkowski} \notag\\
\Sfrw &\equiv \mbox{von Neumann entropy in FRW entangled with Minkowski}\notag
\end{align}
Let $I$ be a subregion in copy 1 and $R$ be a subregion in copy 2. The matter entanglement $\Sflat(I \cup R)$ across the thermofield double has been studied in many papers \cite{1303.1080, 1305.7244, 1311.1200, 1608.05101, 1612.00082, 1803.10244, 1912.11024, 1306.0622}. The details depend on the specific choice of matter, but the general picture is independent of the matter and is easy to understand. In the thermofield double state, the entanglement is local on the scale $\beta_0$. That is,  degrees of freedom near the point $(\eta,x)$ in copy 1 are maximally entangled with degrees of freedom near the same point $(\eta, x)$ in copy 2,  up to a smearing on the scale $\beta_0$.\footnote{
We are choosing the orientation of the time coordinates such that $\eta \to \eta + \delta \eta$, $\bar{\eta} \to \bar{\eta}+\delta \eta$ is a symmetry of the TFD. That is, we evolve \eqref{tfd} under $e^{iH_1\eta_1 - iH_2 \eta_2}$. This is the opposite of the convention in \cite{1303.1080}.
}
This can be seen from the Euclidean path integral used to prepare the thermofield double state.\footnote{We are making the standard assumption that thermal states are gapped up to hydrodynamic modes. This folk theorem may have interesting counterexamples.}

Because entanglement is local across the thermofield double, it follows that if $I$ and $R$ have the same coordinate labels, then $\Sflat(I \cup R)$ is small --- it does not scale with volume.  

Now consider what happens as we deform $I$. If we deform it with $\eta$ fixed, i.e.  $\eta_I= \eta_R$, there is a contribution to the entropy proportional to the non-overlapping volume between $I$ and $R$, because we are no longer keeping all of the entangled partners of the matter in $R$. Thus up to subextensive corrections, 
\be\label{smvol}
\Sflat(I \cup R) \approx s_{\rm th} | V_R - \tV_I |   \qquad \mbox{(equal time)}
\ee
where
 $s_{\rm th}$ is the thermal entropy density.\footnote{This formula assumes that $I \subset R' $ or $R' \subset I$, with $R'$ the image of $R$ in the other copy of Minkowski. Otherwise $|V_R - \tV_I|$ is replaced by the vol$((R'\backslash I) \cup (I \backslash R'))$. Throughout this subsection we will only write the extensive part of the entropy; all of the formulas also have UV-divergent contributions from the boundaries.} The tilde on $\tV_I$ is to emphasize that this is the volume as calculated in flat spacetime, i.e. the comoving volume in FRW.

Now instead of deforming $I$ in the spatial direction, suppose we translate it in time. That is, we take $I$ to have the same spatial domain as $R$, but at $\eta_I \neq \eta_R$. Under time evolution, some of the entangled partners of the matter in $R$ will exit $I$ through its boundary. This leads to a linear-in-time increase in the entropy,
\be\label{smt}
\Sflat(I \cup R) \approx s_{\rm th} v_E |\eta_I - \eta_R| \wArea(\p I) 
\qquad \mbox{(equal space)} \ .
\ee
Here $v_E$ is a proportionality constant known as the entanglement velocity \cite{1303.1080, 1305.7244, 1311.1200} which is constrained by general arguments to the range $0 < v_E \leq 1$ \cite{1509.05044, 1512.02695}. Again the tilde means `comoving.' This formula holds for time scales larger than the thermal wavelength, 
\be\label{etadiff}
|\eta_I - \eta_R| \gtrsim \beta_0 \ ,
\ee
but much less than the extrinsic curvature scale of region $I$ or $R$.\footnote{
The meaning of `$\gtrsim$' in \eqref{etadiff} requires more explanation. For the most part, we will be interested in derivatives of $\Sflat$. In this case, we expect \eqref{smt} to be accurate for $|\eta_I - \eta_R|$ larger than some $O(1)$ number times $\beta_0$. It is not necessary to take $|\eta_I - \eta_R|\gg \beta_0$. For example in 2d we can see from the explicit formula \eqref{sia1} that the derivative of \eqref{smt} is accurate to one part in $10^4$ already for $|\eta_I - \eta_R| = \beta_0$. However there is also a constant shift in the exact formula compared to \eqref{smt}, so the actual value is only accurate for $|\eta_I - \eta_R| \gg \beta_0$.}

\begin{figure}
\begin{center}
\includegraphics[scale=1]{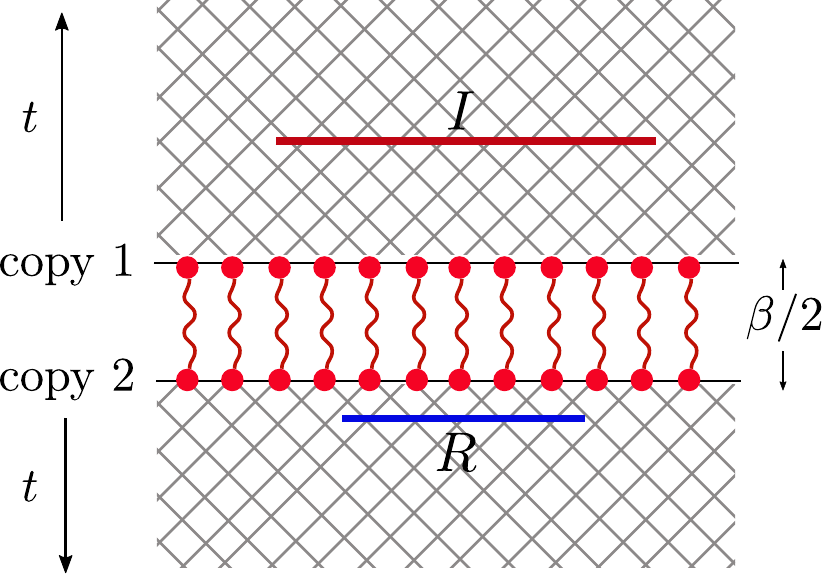}
\end{center}
\caption{\small Quasiparticle picture of entanglement across the thermofield double.  \label{fig:quasiparticles}}
\end{figure}

If $I$ is deformed in both space and time, both effects increase the entropy. We will consider some examples below.  The general conclusion is that $\Sflat(I \cup R)$ is minimized when $I$ and $R$ have the same coordinate values in both space and time, and if we deform $I$ slightly, then it takes the form
\be\label{flatdef}
\Sflat(I \cup R) \approx s_{\rm th} f(\eta_I - \eta_R, r_I - r_R) \wArea(\p I) \qquad \mbox{(general small deformation)} \ .
\ee
The formulas \eqref{smt} and \eqref{etadiff} determine $f(\delta \eta,0)$ and $f(0, \delta r)$ for $|\delta \eta|$, $|\delta r|$ much larger than the scale $\beta_0$ set by the temperature but much smaller than the sizes of the regions. There is also a universal formula in the limit of $|\delta \eta|, |\delta r|$ much less than $\beta_0$ that we will discuss below.

These results can be understood qualitatively in terms of a quasiparticle picture for entanglement spreading, first introduced by Cardy and Calabrese in the study of global quenches \cite{cond-mat/0601225, 0708.3750} and illustrated in figure \ref{fig:quasiparticles}. The state is prepared by Euclidean evolution by $\beta_0/2$, which creates short-range entanglement between pairs of quasiparticles on opposite sides of the thermofield double. Under Lorentzian time evolution, the entangled quasiparticles spread out along the gray lines. The entanglement entropy $\Sflat(I \cup R)$ is estimated by counting the number of quasiparticles in $I \cup R$ whose partners are not in $I \cup R$.

Another general property of the function $f$ that will be useful is that it cannot change too rapidly. This is intuitively clear, because $f$ is like a hydrodynamic entropy, which can change under spatial deformations by at most $s_{\rm th}\delta \mbox{Vol}$ and similarly under timelike deformations. This intuition can be formalized using the monotonicity of relative entropy, which leads to the bounds \cite{1512.02695}
\be\label{fderivs}
| \p_{\eta_I} f(\eta_I - \eta_R, r_I - r_R) | \leq 1 , \quad | \p_{r_I} f(\eta_I - \eta_R, r_I - r_R) | \leq 1 \ .
\ee
We review the derivation for a spacelike derivative in appendix \ref{app:fderiv} and refer the reader to \cite{1512.02695} for the timelike case. 

For holographic matter, the function $f$ is known explicitly for symmetric regions. In appendix \ref{app:cft2tfd} we derive exact formulas for a holographic 2d CFT, which match this general discussion. In that case, $v_E=1$ and the function $f$ appearing in \eqref{flatdef} is
\be
f_{2d}(\delta \eta, \delta r) = \frac{2\pi c}{3\beta_0} \max\left( |\delta \eta|, \ |\delta r |\right) \ .
\ee
In higher dimensions, $f$ for holographic matter is derived in \cite{1803.10244}. 

\subsection{Turning on the scale factor}

So far this discussion has ignored the FRW scale factor. We can incorporate it by doing a conformal transformation.\footnote{See \cite{1210.7244} for a discussion of entanglement entropy in de Sitter space, including the more difficult case of massive fields.} The effect is to rescale the UV divergence in the matter entropy,
\be
\epsilon_{\rm uv}  \to \frac{ \epsilon_{\rm uv} }{a(\eta)} \ .
\ee
In odd dimensions, this has no effect on the generalized entropy, because the divergences in the matter entropy are absorbed into counterterms and the renormalization of Newton's constant, as discussed in section \ref{ss:condition1}.\footnote{We assume all regions have a smooth boundary so there are no corner contributions.}  In even dimensions, depending on the shape of the region, there can be a logarithmic divergence $C \log \epsilon_{\rm uv}$ controlled by the $a$-type conformal anomaly.  Therefore the effect of the scale factor is 
\be\label{sfrwr}
\Sfrw(I \cup R)  = \Sflat(I \cup R)  + C \log a(\eta_I) +\cdots \ ,
\ee
where the dots are absorbed into the area term in the definition of the generalized entropy. For an interval in two dimensions, $C = \frac{c}{3}$ with $c$ the central charge. In four dimensions, for a half-space $C=0$, and for a sphere 
\be\label{sphereweyl}
C =  -4A
\ee
with $A$ the Euler-type Weyl anomaly\cite{hep-th/0603001, 0802.3117}.
%0905.0932, 1008.4314, 1007.1813, 1011.5819, 1102.0440

Combining \eqref{sfrwr} with \eqref{flatdef} we have
\be\label{finalsfrw}
\Sred(I \cup R) \approx s_{\rm th}f(\eta_I - \eta_R, r_I - r_R) \wArea(\p I)  + C \log a(\eta_I)  + \Sdiv(\p R) \ .
\ee
Recall that $\Sred$ is defined in \eqref{defsred} by subtracting the UV divergence at $\p I$ but not at $\p R$, so in this formula we have included all of the extensive terms and UV divergences in $\Sred$. The area here is the comoving area, and $s_{\rm th}$ is the comoving entropy density, i.e. entropy per coordinate volume in FRW. This formula holds for small deformations away from $I=R$, on scales much larger than $\beta_0$ but small compared to the size of the regions, the extrinsic curvature of the regions, or the spacetime curvature.

\section{No islands in radiation-dominated FRW}\label{s:radiation-only}
Consider a four-dimensional radiation-only FRW cosmology. The energy density and scale factor are
\be
\epsilon = \frac{\epsilon_0}{a^4} \ , \quad
a = \frac{\eta}{\eta_0} ,  \quad \eta_0 = \sqrt{ \frac{3}{8\pi \epsilon_0}} \ .
\ee
In conformal coordinates, $T_{\mu\nu} = a^2 \epsilon \mbox{diag}(1, \frac{1}{3}, \frac{1}{3}, \frac{1}{3})$. This state of the CFT is conformally related to a finite-temperature state in Minkowski spacetime with energy density
\be
\epsilon_0 = c_{\rm th}T_0^4 
\ee
and thermal entropy density
\be
s_{\rm th} = \frac{4}{3} c_{\rm th} T_0^3 \ .
\ee
The constant $c_{\rm th}$ is roughly proportional to the number of degrees of freedom, and $T_0 = \beta_0^{-1}$ is a constant parameter that corresponds to the temperature of the state in flat spacetime, before the Weyl transformation to FRW. $T_0$ is equal to the physical temperature in FRW at $\eta = \eta_0$.

\subsection{Application of the general constraints}
 
\begin{figure}
\begin{center}
\includegraphics[scale=0.7]{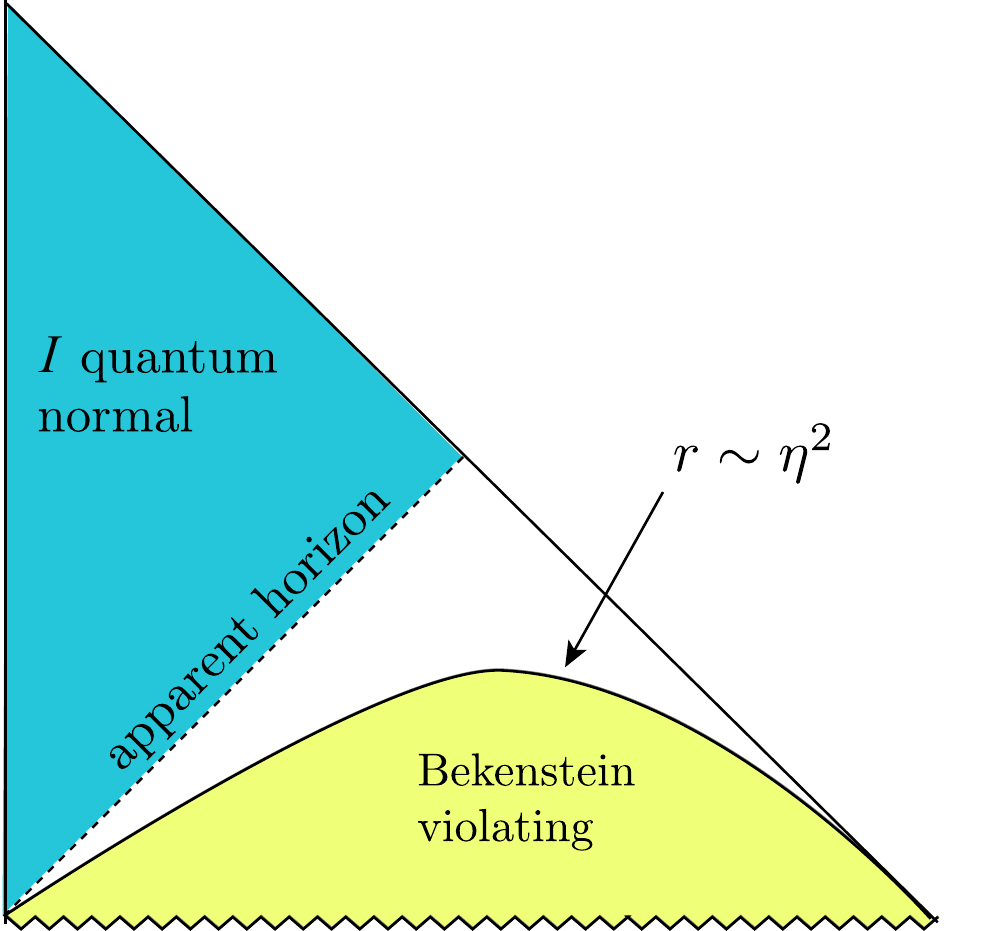}
\end{center}
\caption{
Regions of the radiation-only FRW cosmology, defined for a spherical region $I$. The Bekenstein-violating region does not overlap with the quantum normal region for $I$, and there is no quantum normal region for $G$, so there cannot be spherical islands.
\label{fig:radiation-only}}
\end{figure}

Let $I$ be a spherical region of comoving radius $r_I$ at time $\eta_I$. The matter entropy is dominated by the extensive, thermodynamic contribution, so the generalized entropy associated to this region is
\begin{align}
S_{\rm gen}(I) &\approx \frac{\Area(\p I)}{4} + s_{\rm th}\tVol(I) \label{sgenIfrw} \\
&= \pi a_I^2 r_I^2 + \frac{4\pi}{3}s_{\rm th} r_I^3 \ .  %number both lines
\end{align}
Here $\tVol$ is the comoving volume and $a_I := a(\eta_I)$. 
Comparing this to the area term, we see that the Bekenstein area bound \eqref{matterbound2} is violated for
\be\label{radiationBek}
r_I \gtrsim \frac{3\pi}{2} T_0 (\eta_I)^2 \ .
\ee
This is the first condition.
Now we will identify the quantum normal regions. Recall from section \ref{s:generalconditions} that this is the region in which $S_{\rm gen}$ increases under forward-directed outward deformations, and decreases under forward-directed inward deformations. That is, the quantum normal region for $I$ is the region satisfying
\be\label{normali}
\pm(\p_{\eta_I} \pm \p_{r_I}) S_{\rm gen}(I) \geq 0  \ .
\ee
The outgoing condition $(+)$ is always satisfied. The ingoing condition $(-)$ is satisfied in the region
\be\label{normal1}
r_I \leq r_{QAH} = \begin{cases}
 \frac{\pi T_0 (\eta_I)^2}{\pi T_0 \eta_I -1}  & \eta_I > \frac{1}{\pi T_0} \\
 \infty & \eta_I < \frac{1}{\pi T_0}
 \end{cases}
 \ee
The time $\eta = 1/(\pi T_0)$ when the quantum apparent horizon goes to infinity is a Planck distance from the big bang singularity so the semiclassical theory does not apply, and it cannot be trusted. To see this, note that the proper time elapsed from $\eta=0$ to $\eta = 1/(\pi T_0)$ is $t = \sqrt{ 2 c_{\rm th} \over 3\pi^3}$, and we are working in Planck units.
In the semiclassical regime with $T_0 \eta_I \gg 1$, the quantum normal region for $I$ is therefore simply
\be\label{radiationNormal}
r_I \leq \eta_I \ .
\ee
This is the same as the classical normal region. The quantum normal region for $I$ and the Bekenstein-violating region are shown in figure \ref{fig:radiation-only}. 

The third condition states that region $G$ is also quantum normal. (See figure \ref{fig:frwisland} for the definition of region $G$.) The generalized entropy of $G$ is
\begin{align}
S_{\rm gen}(G) &= \frac{\Area(\p I)}{4}  - s_{\rm th} \tVol(I) + \mbox{const.} \label{sgenGfrw}\\
&=\pi a_I^2r_I^2 - \frac{4\pi}{3}s_{\rm th}r_I^3 + \mbox{const.} %number both lines
\end{align}
The quantum normal condition is
\be
\pm (\p_{\eta_I} \mp \p_{r_I})S_{\rm gen}(G) \geq 0 \ .
\ee
This requires the outgoing expansion to be positive and the ingoing expansion to be negative. The signs differ from \eqref{normali} because the definition of `outgoing' is opposite for region $G$. Satisfying the ingoing condition requires $\eta_I < \beta_0/\pi$, which is outside the semiclassical regime.

In section \ref{s:generalconditions} we showed that region $I$ can only be an island if all three conditions are satisfied. Clearly this is impossible in the semiclassical regime, because condition 1 and condition 2 have no overlap, and condition 3 is never satisfied. Therefore the four-dimensional radiation-only FRW cosmology has no spherically symmetric islands.

\newcommand{\sgn}{\mbox{sgn}}
\subsection{Explicit check}
Of course we can also look for islands by extremizing the generalized entropy explicitly. Let $R$ be a spherical region in the auxiliary Minkowski spacetime that purifies the matter state in FRW. It is defined at time $\eta_R$ and has radius $r_R$. We will look for an island in FRW for which $I$ is a small deformation of $R$, i.e. $\eta_I = \eta_R + \delta \eta$ and $r_I = r_R + \delta r$ with $|\delta r|, |\delta \eta| \ll  r_R$. Using \eqref{sphereweyl}-\eqref{finalsfrw}, the generalized entropy that appears in the island formula takes the form
\begin{align}\label{radgen}
S_{\rm gen}(I \cup R) &= \frac{1}{4}\Area(\p I) + \Sred(I \cup R) \\ 
&\approx \pi a_I^2 r_I^2 + 4\pi s_{\rm th}f(\delta \eta, \delta r) r_I^2 - 4A \log a_I  + \Sdiv(\p R) \ . \notag
\end{align}
The function $f(\delta \eta, \delta r)$ is complicated (unknown in general), but the entanglement speed limit derived in \cite{1512.02695} and reviewed around equation \eqref{fderivs} requires
\be
|\p_{\eta_I}f(\delta \eta, \delta r) | \leq 1 \ , \quad
|\p_{r_I}f(\delta \eta, \delta r) | \leq 1  \ .
\ee
It follows that the extremality condition $\p_{\eta_I}S_{\rm gen} =0$ cannot be satisfied, except possibly in the Planck regime, $\eta_I \lesssim \beta_0$, because it is only in this regime that the matter derivative can compete with the derivative of the area term. Therefore there are no islands in the semiclassical regime.\footnote{Formally, using our formulas for the entropy leads to an island near the FRW singularity, $\eta_I \sim 0$. It is a minimum of the generalized entropy in the time direction so it entails a formal violation of quantum focusing. However this is outside the validity of the semiclassical theory.}

\subsection{Positive cosmological constant}

\begin{figure}
\begin{center}
\includegraphics[scale=0.5]{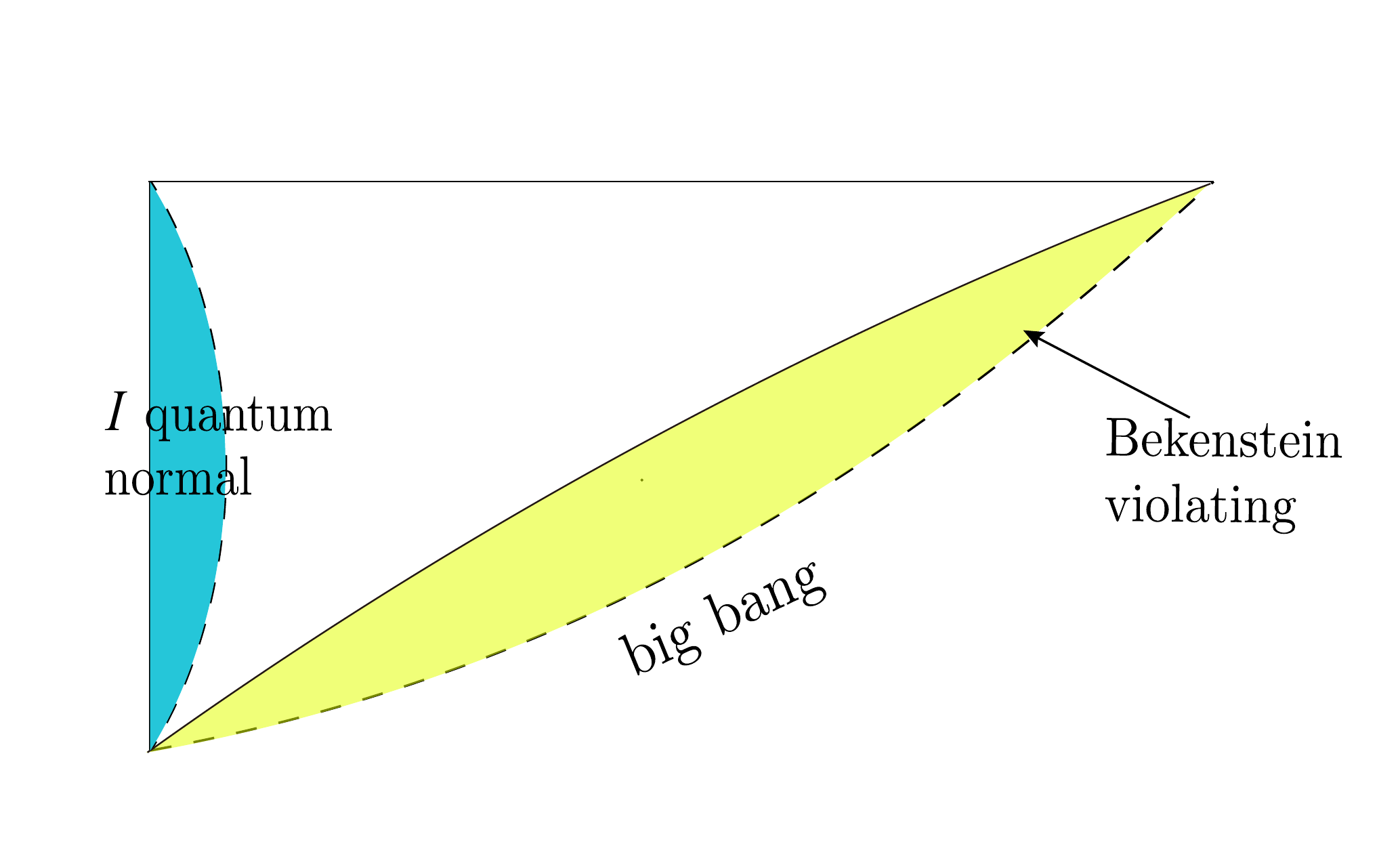}
\end{center}
\caption{
Regions of the FRW cosmology with radiation and positive vacuum energy. The Bekenstein-violating region does not overlap with the quantum normal region for $I$, and there is no quantum normal region for $G$, so there cannot be spherical islands.
\label{fig:positiveCC}}
\end{figure}

Now we turn on a positive cosmological constant $\Lambda>0$.  In the FRW coordinates \eqref{frwcoords}, the Friedmann equation is 
\be\label{ccfr}
3 \left( \frac{\dot{a}}{a} \right)^2 = \frac{8 \pi  \epsilon_0}{a^4} + \Lambda \ .
\ee
The solution is 
\be
a(t)=a_0 \sqrt{\sinh{\frac{\pi t}{2t_m} }}
\ee
where
\be
a_0 = \left( \frac{8 \pi \epsilon_0}{\Lambda} \right)^{1/4} , \qquad t_m = \frac{\pi}{4} \sqrt{ \frac{3}{\Lambda}} \ .
\ee
The big bang singularity is at $t=0$. As in the discussion above, the Bekenstein area bound  \eqref{matterbound2}  is violated for
\be\label{radiationCCbek}
r_I \gtrsim r_{Bek}=\frac{3 a^2}{4 s_{th}}\,.
\ee
The outgoing quantum normal condition for $I$ is always satisfied, and the ingoing condition for $I$ is satisfied (in the semiclassical regime $t \gg \ell_P$) when 
\be\label{radiationCCqah}
r_I \leq r_{QAH} \approx r_{AH}= \frac{1}{da/dt} \ .
\ee
These regions are shown in figure \ref{fig:positiveCC}.
The ingoing quantum normal condition for $G$ is again never satisfied. We conclude that there are no spherically symmetric islands in four-dimensional FRW cosmology with radiation and a positive cosmological constant.

\section{Islands in recollapsing FRW}\label{s:recollapse}

We will now consider a four-dimensional FRW cosmology with radiation and a negative cosmological constant $\Lambda<0$. Solving the Friedman equation \eqref{ccfr}, the solution for the scale factor in FRW coordinates is

\be
a(t) =a_0 \sqrt{ \cos \frac{\pi t}{2t_m} }
\ee
where again
\be
a_0 = \left( \frac{8 \pi \epsilon_0}{|\Lambda|} \right)^{1/4} , \qquad t_m = \frac{\pi}{4} \sqrt{ \frac{3}{|\Lambda|}} \ .
\ee
The big bang is at $t = -t_m$. The spacetime begins to recollapse at the turning point $t=0$, and there is a crunch singularity at $t = t_m$.

\subsection{Bekenstein area bound and the quantum normal regions}

It is useful to restore units temporarily to see how things scale with the Planck length $\ell_P \sim \sqrt{\hbar G_N}$. The crunch time scales as
\be
t_m \sim \frac{1}{\sqrt{|\Lambda|}} \ .
\ee
The maximal scale factor is
\be
a_0 \sim (G_N \epsilon_0 / |\Lambda|)^{1/4} \ .
\ee
This is dimensionless and assumed to be $O(\ell_P^0)$, so
\be
\frac{\hbar G_N}{\Lambda \beta_0^4} \sim O(\ell_P^0)\,,
\ee
where $\beta_0$ is the thermal wavelength of the matter. Therefore $t_m$ scales as $\beta_0^2/\ell_P$. The entropy density is $s_{\rm th} \sim 1/\beta_0^3$ independent of $\ell_P$. So in understanding the semiclassical regime we should scale the island time as $t_I \sim \beta_0$ or $t_I \sim \beta_0^2/\ell_P$, and hold everything else fixed.

\begin{figure}
\begin{center}
\includegraphics[scale=0.7]{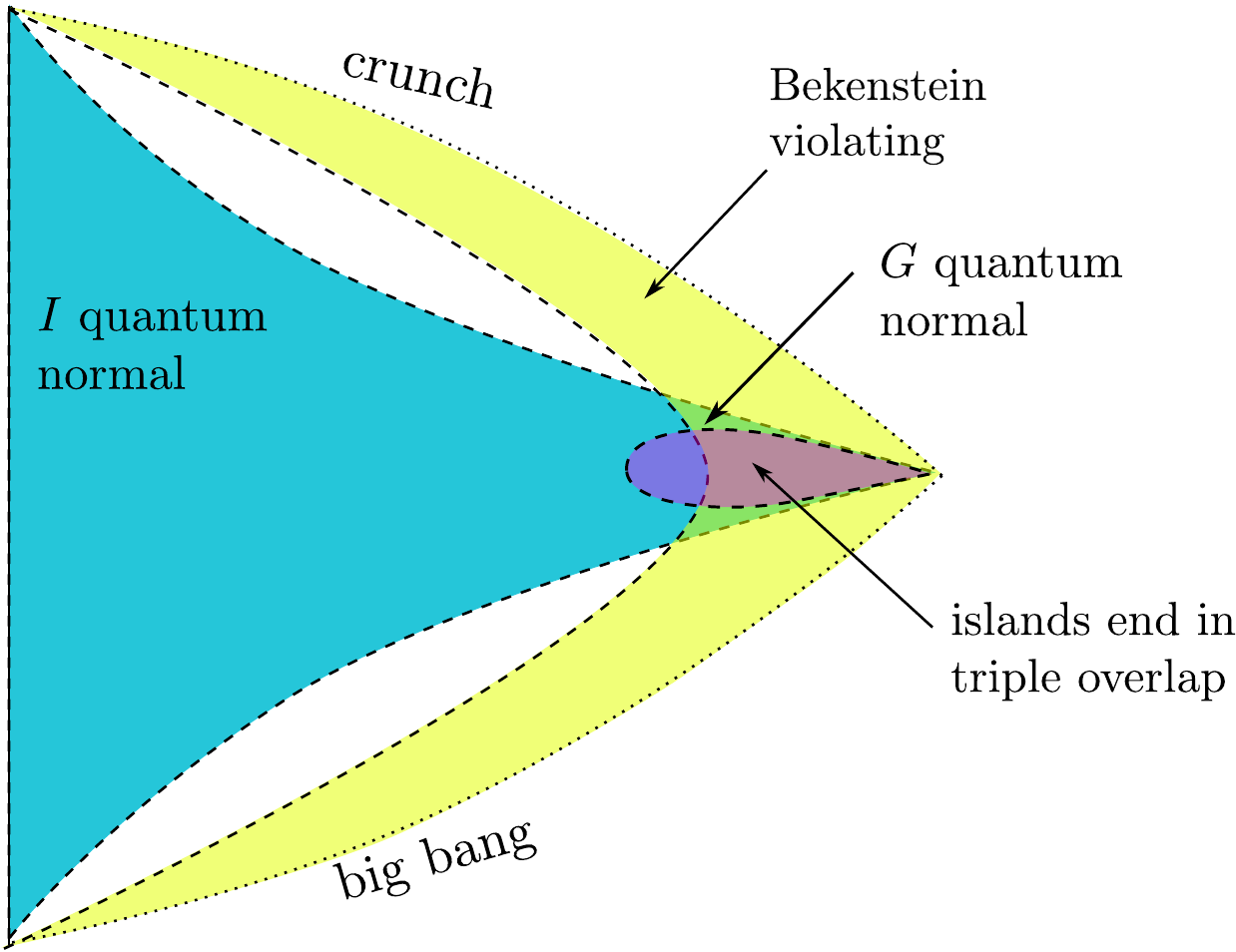}
\end{center}
\caption{Recollapsing FRW cosmology with radiation and a negative CC.\label{fig:recollapse-regions}}
\end{figure}

We will now return to natural units. The generalized entropy of a spherical region $I$ is again given by \eqref{sgenIfrw} with this new scale factor. The Bekenstein area bound is violated for
\begin{align}
r_I \gtrsim  r_{\rm Bek} =  \frac{3a_0^2  }{4 s_{\rm th} } \cos \frac{\pi t_I}{2 t_m} \ .
\end{align}
The quantum normal region for $I$ is defined by the two inequalities
\be
(\pm a_I \p_t +  \p_r) S_{\rm gen}(I) \geq 0 \ .
\ee
These are equivalent to the condition
\be\label{qah}
r_I\left[2s_{th}t_m - \frac{\pi}{4}a_0^3\sin \frac{\pi |t_I|}{2t_m}  \sqrt{\cos \frac{\pi t_I}{2t_m}}\right]
 + a_0^2 t_m \cos \frac{\pi t_I}{2t_m} \geq 0 \ . 
 \ee
We restrict our attention to the regime of semiclassical gravity, away from the singularities. Far away from the turning point $t_I=0$, the matter term drops out because it is suppressed by $\ell_P$ when we restore units. In this case the quantum normal region agrees with the classical normal region, which is 
\be\label{rAH}
r \leq r_{AH} = \frac{4t_m}{\pi a_0} \frac{\sqrt{ \cos \frac{\pi t_I}{2t_m} }}{ \sin \frac{\pi |t_I|}{2t_m} } \ .
\ee 
Near the turning point, the matter term in \eqref{qah} is important. In this regime we can expand the trigonometric functions in \eqref{qah}. Define 
\be
t_1 = \frac{16 s_{th} t_m^2 }{\pi^2 a_0^3}  = \frac{ a_0 \beta_0}{2\pi}.
\ee
We find the quantum normal region for $I$ in the regime $t_I \ll t_m$ is 
\be\label{rQAH}
r_I \leq r_{QAH} = \begin{cases}
\infty & |t_I | \leq t_1   \\
\frac{8t_m^2}{\pi^2a_0(|t_I| - t_1)}  & |t_I| > t_1
\end{cases}
\ee
Note that using the scalings above, $t_1$ is a macroscopic (non-Planckian) timescale of order the thermal wavelength. For $|t_I| \gg t_1$, the quantum apparent horizon $r_{QAH}$ approaches the classical apparent horizon $r_{AH}$. The quantum apparent horizon hits the boundary of the Bekenstein-violating region, $r_{Bek} = r_{QAH}$, at $t = \frac{5}{3}t_1 = \frac{5}{6\pi}\beta$, where $\beta = a_0 \b_0$ is the physical inverse temperature at the turning point.

The generalized entropy of region $G$ is given by \eqref{sgenGfrw}, and the corresponding quantum normal condition is
\be
(\pm a_I \p_t - \p_r) S_{\rm gen}(G) \geq 0 \ .
\ee
This condition is satisfied only in the small teardrop region on the Penrose diagram in figure \ref{fig:recollapse-regions}. Intuitively, the reason the condition is hard to satisfy is that $G$ is the outside of a sphere. Classically, the outside of a sphere in a weakly curved spacetime is anti-normal, so a large matter contribution is required to overcome the area term.

Islands can exist only in the triple overlap where $I$ is Bekenstein-violating and quantum normal, and $G$ is quantum normal. This requires $I$ to be a large region near the turning point. The situation is summarized in figure \ref{fig:recollapse-regions}.

\subsection{Islands for $|t_R| \lesssim \beta$}

In the auxiliary region, it is convenient to define a rescaled time coordinate
\be
t_R \equiv a_0 \eta_R  \ .
\ee
Any spherical islands in the radiation+negative CC cosmology must have their boundary in the overlap of the Bekenstein-violating region with the quantum normal region. We have determined that this region is limited to the corner of the Penrose diagram near spatial infinity, with $|t_I| \lesssim \beta$ and $r_I \gtrsim \frac{3a_0^2}{4s_{\rm th}}$. We will now find that there are indeed islands in this regime if we calculate the entropy of a region $R$ in the auxiliary system with $|t_R| \lesssim \beta$ and $r_R \gtrsim \frac{3a_0^2}{4s_{\rm th}}$.

Using \eqref{finalsfrw}, the generalized entropy that appears in the island formula again takes the form
\begin{align}\label{sgenREC}
S_{\rm gen}(I \cup R) &= \frac{1}{4}\Area(\p I) + \Sred(I \cup R) \\ 
&\approx \pi a_I^2 r_I^2 + 4\pi s_{\rm th}f(\delta \eta, \delta r) r_I^2 - 4A \log a_I  + \Sdiv(\p R) \ .\notag
\end{align}
with
\be
\delta \eta = \eta_I - \eta_R , \quad \delta r = r_I - r_R \ .
\ee
The behavior of $f(\delta \eta, \delta r)$ is described in section 4, and the relation between conformal time $\eta$ and cosmological time $t$ is given in \eqref{conformalcoords}. The last term is the UV divergence from region $R$, which does not affect the extremization.

To find an island, we fix the region $R$ and look for extrema of $S_{\rm gen}$. Assume $t_R \lesssim \beta$ and $r_R \gtrsim \frac{3 a_0^2}{4 s_{\rm th}}$. It is easy to see that there is an extremum near $t_I \approx t_R$, $r_I \approx r_R$. We will check this self-consistently by first setting $\delta \eta = 0$ and looking at $S_{\rm gen}$ as a function of $r_I$:
\be
S_{\rm gen}(I \cup R) \approx \pi a_0^2 r_I^2 + 4\pi s_{\rm th}r_I^2 f(0, r_I - r_R) + \mbox{const.} 
\ee 
Using \eqref{smvol} to evaluate $f$, and $r_R \gtrsim r_{Bek}$, this has a minimum near $r_I \approx r_R$ up to corrections of $O(\beta)$. 

Now we set $r_I = r_R$, and consider the behavior of $S_{\rm gen}$ as a function of $t_I$. 
We are looking for islands with $|t_I| \ll t_m$, so we can expand around the turning point, 
\be
\eta_I \approx \frac{t_I}{a_0} , \quad a_I\approx a_0(1 - \frac{|\Lambda|}{3}t_I^2)   \ .
\ee
In the regime $ |t_I| \ll t_m$, with $r_I = r_R$, the entropy is therefore
\be
S_{\rm gen}(I \cup R) \approx \mbox{const.} - \frac{2\pi}{3}  a_0^2 |\Lambda| r_R^2 t_I^2 + 4 \pi s_{\rm th}r_R^2  f(\frac{t_I-t_R}{a_0},0)    \ .
\ee
We have dropped the much smaller term from the anomaly, $\frac{4}{3}A |\Lambda| t_I^2$. If additionally $|t_I| \gtrsim \beta$, then we can also use \eqref{smt} to evaluate $f$, and we find
\be\label{soft2}
S_{\rm gen}(I \cup R) \approx \mbox{const.} - \frac{2\pi}{3}  a_0^2 |\Lambda| r_R^2 t_I^2 + 4\pi s_{\rm th} r_R^2 v_E \frac{|t_I-t_R|}{a_0} \ .
\ee
This function is symmetric about the origin, and it is increasing when $t_I \lesssim -\beta$, and decreasing when $t_I \gtrsim \beta$. Therefore there is a maximum around $t_I \approx 0$, up to corrections of $O(\beta)$.

The conclusion is that when $R$ is a large enough region near the point of maximal scale factor, there is an island with $I \approx R$.  This dominates over the trivial island for large regions, because the island entropy does not grow with volume. The entropy of region $R$ without including an island is simply its thermal entropy, which does grow with volume.

\subsubsection{Subleading analysis and the QNEC bound}\label{ss:qnecbound}
The approximations used thus far can only pinpoint the location of the island to an accuracy of $O(\beta)$. In some cases we can do better. For these purposes let us assume $t_R = 0$, $r_R > r_{Bek}$. The function written in \eqref{soft2} actually has two local maxima at 
$t_I=\pm  v_E \beta/(2\pi)$ and a local minimum at $t_I=t_R=0$. However as all three of these extrema have $t_I = O(\beta)$, the approximations used to derive \eqref{soft2} are inaccurate in this regime. Indeed, from  \cite{1912.02799} and the discussion in section \ref{ss:maximin} we expect the island to be a maximum of $S_{\rm gen}$ in the time direction, so the minimum should disappear under closer scrutiny. We will now check this, and find a nice agreement with an entropy bound derived recently by Mezei and Virrueta  using the quantum null energy condition (QNEC) \cite{1909.00919}.

In the regime $|t_I| \ll \beta$, the function $f$ is known to be quadratic in time \cite{cond-mat/0503393, 1909.00919, 1305.7244, 1311.1200},
\be
s_{\rm th} f(\delta \eta, 0) \approx b_f (\delta \eta)^2  \qquad (|\delta \eta| \ll \beta_0) \ .
\ee
Using this approximation, we find that for $t_I \ll \beta$, the time dependence of the generalized entropy, after setting $r_R = r_I$, is 
\be
S_{\rm gen}(I \cup R) \approx \mbox{const.}  - \frac{2\pi}{3} r_R^2 a_0^2 |\Lambda|\left(1 - \frac{6 b_f}{a_0^4 |\Lambda|}\right) t_I^2 \ .
\ee
Thus $t_I = 0$ is a maximum of the generalized entropy if 
\be
b_f < \frac{a_0^4 |\Lambda|}{6} = \frac{4}{3}\pi \epsilon_0 \ .
\ee
The QNEC bound \cite{1909.00919} for a 4d CFT is $b_f\leq \frac{4}{3}\pi \epsilon_0$. If the QNEC bound were violated, then the island would be a minimum under time deformations, and quantum focusing would also be violated. This fits together nicely, since the QNEC was originally motivated by quantum focusing. In a holographic theory the QNEC bound is saturated. Then the quadratic correction to $S_{\rm gen}$ vanishes so we would need to go to higher order to resolve the extremum.

\subsection{Islands for $|t_R| \gtrsim \beta$}
So far we have only analyzed the case where region $R$ is chosen to be near the turning point, $t_R \approx 0$. We will now ask what happens as we increase $t_R$. The matter contribution is quite complicated in general, so we will not attempt a general analysis. We will just ask what happens to the island we have found at the turning point if we increase $t_R$, assuming $t_R \ll r_R$. The spacetime is symmetric in time, so without loss of generality we choose $t_R > 0$. 

For concreteness we will assume the holographic formula for the matter entropy function, $f$. Although the entropy is different in other theories, the general shape is similar in any CFT, so the conclusions about the island would also be qualitatively similar. 

\begin{figure}[t]
\begin{center}
\includegraphics{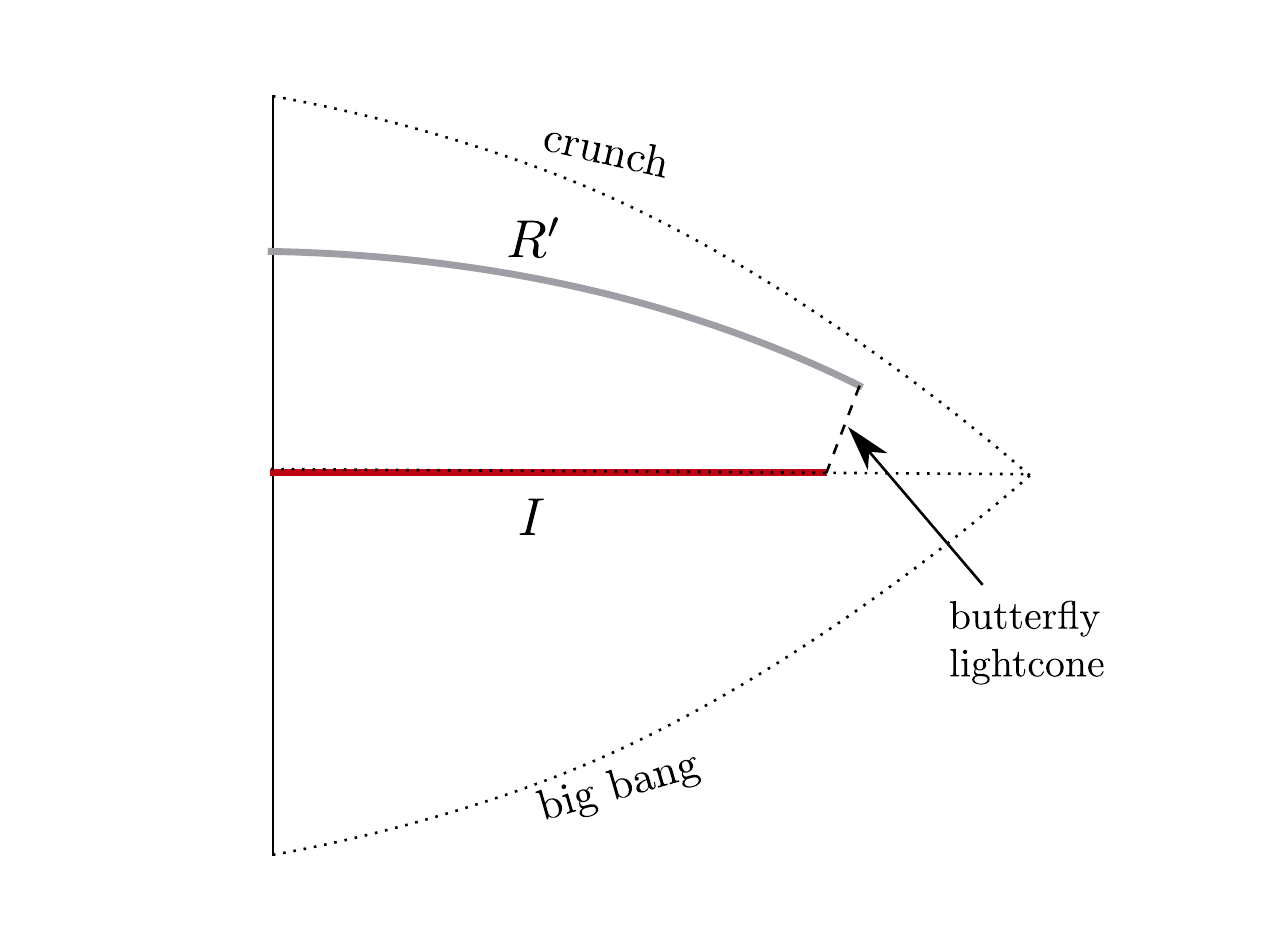}
\end{center}
\caption{Island for a region $R$ that is not at $\eta_R =0$. This is an example where $r_I$ is at the lower end of the range \eqref{rangeR}. 
\label{fig:shiftedisland}}
\end{figure}

We will first minimize $S_{\rm gen}$ as a function of $r_I$. We can neglect the anomaly term, and we know from the general conditions that the island lies near the turning point, so the function we are extremizing is
\be\label{bbmin}
S_{\rm gen}(I \cup R)   =  \mbox{const.} - \frac{2\pi}{3}  a_0^2 |\Lambda| r_I^2 t_I^2 + 4 \pi r_I^2 s_{\rm th} f(\eta_I - \eta_R, r_I - r_R) \ .
\ee
In the regime $ |\delta \eta|, |\delta r| \ll r_R$, the matter entropy in a holographic CFT has \cite{1303.1080, 1305.7244, 1311.1200, 1803.10244}
\be\label{fstrip}
  f(\eta_I-\eta_R, r_I-r_R) =
  \begin{cases}
                                   |\delta \eta| \frac{v_E}{\left(1-(\frac{\delta r}{\delta \eta})^2\right)^{1/4}} & \text{if $|\delta r| \leq v_B |\delta \eta|$} \\
                                   |\delta r| & \text{if $|\delta r|>v_B |\delta \eta|$} 
  \end{cases}
\ee
where $v_B = \sqrt{\frac{2}{3}}$ is the butterfly velocity \cite{1306.0622} and $v_E = \frac{\sqrt{2}}{3^{3/4}}$ is the entanglement velocity in four dimensions \cite{1303.1080}. Here as  $ |\delta \eta|, |\delta r| \ll r_R$, we are using the exact result for a strip.

We know from the Bekenstein area bound and quantum normal condition that $|\eta_I|$ is small, so we can replace $|\eta_R - \eta_I| \approx \eta_R$. If $r_R$ is large enough, then \eqref{bbmin} has a minimum in the spatial direction that lies in the range
\be\label{rangeR}
r_I \in [r_R - v_B \eta_R, r_R] \ .
\ee
The formula for the minimum is not illuminating so we will not reproduce it. As $\eta_R$ increases, the minimum moves to the lower endpoint, $r_I \approx r_R - v_B \eta_R$.
The time dependence is dominated by the area term for $t_I \gtrsim \beta$, so there is a maximum at $t_I \approx 0$, up to $O(\beta)$ corrections.

Therefore we conclude that as $\eta_R$ is increased, the quantum extremal surface stays at the turning point of the scale factor, but moves to smaller $r$. An example is drawn in figure \ref{fig:shiftedisland}. In the figure, $R'$ is the partner of region $R$ in FRW, i.e., $(r_{R'}, \eta_{R'}) = (r_R, \eta_R)$.

Note that at any fixed time $\eta_R$ in the auxiliary spacetime, there will always be a nontrivial island if we choose $r_R$ large enough. At any fixed $r_R \gtrsim r_{Bek}$, there is a finite range of $\eta_R$ with a nontrivial island; the size of this range is controlled by $r_R - r_{Bek}$, with larger regions having nontrivial islands at larger times.\footnote{We have only analyzed the extremum that appears for $t_R  = 0$, and followed it as we change $t_R$. We have not ruled out the possibility that for some choice of matter sector, $f$ could be such that there are other extrema that appear in other parameter ranges.}

\section{dS$_2$}\label{s:dS2}

\begin{figure}
\begin{center}
\includegraphics[scale=.4]{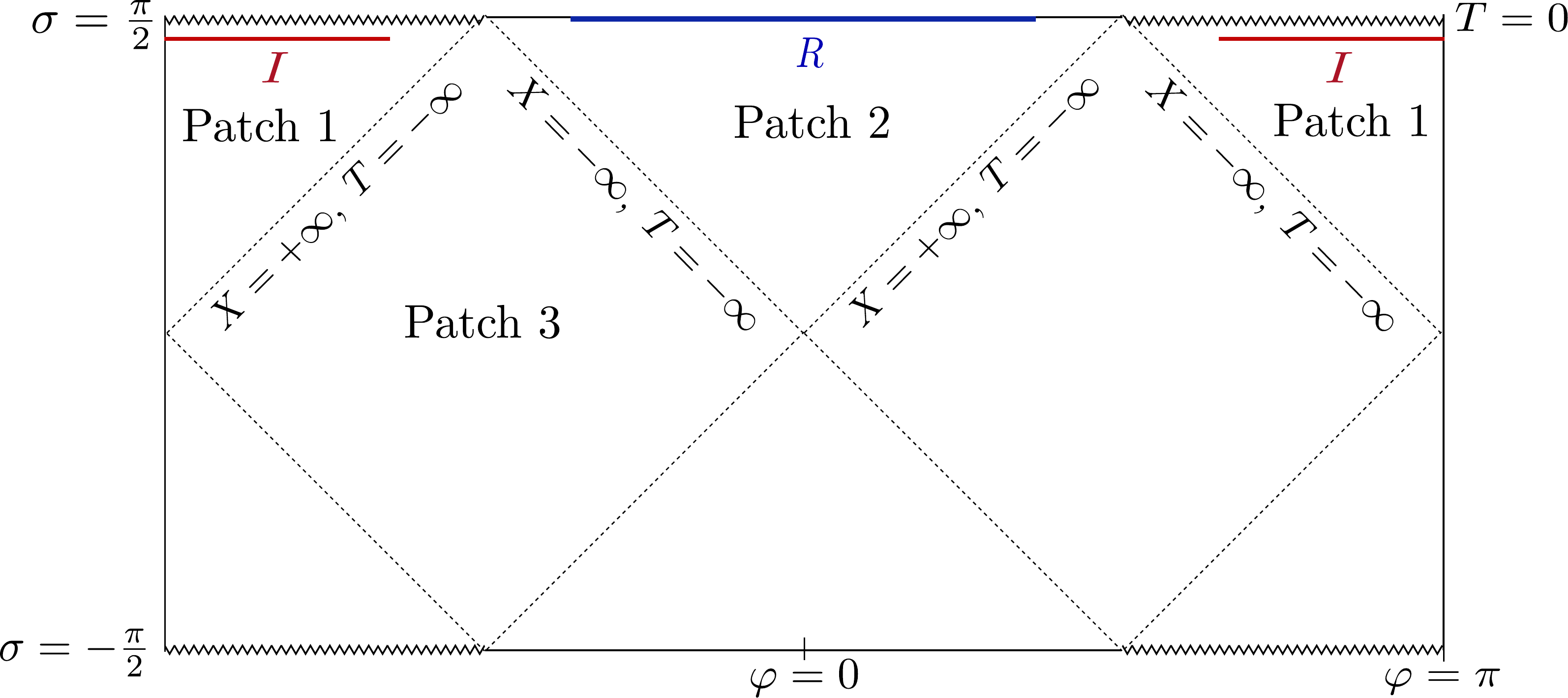}
\end{center}
\caption{Penrose diagram of dS$_2$. Patch 1 and 2 are often called the hyperbolic patches. 
Patch 3 is a static patch. The left and right edges of the diagram are identified. Region $R$ sits near $\mathcal{I}^+$ and has an island region $I$. 
} \label{fig:dsisland}
\end{figure}

In this section we will consider two-dimensional de Sitter spacetime in JT gravity \cite{Teitelboim:1983ux, Jackiw:1984je} coupled to a two-dimensional CFT. The action is
\be
S =  \f{\phi_0}{16\pi G}\int d^2 x \sqrt{-g} R+ \f{1}{16\pi G}\int d^2 x \sqrt{-g}\, \phi(R-2) + S_{\rm CFT}\,,
\ee
where we have left out boundary terms. For details about this model see \cite{Anninos:2017hhn, Anninos:2018svg, Maldacena:2019cbz, Cotler:2019nbi}. The first term (combined with a boundary term) is purely topological and will not play a role in our discussion, except for a constant shift in the entropy. The equations of motion of this theory, with vanishing stress tensor for the CFT, are
\be
R = 2 \,,\qquad (g_{\m\n} \nabla^2 - \nabla_\m \nabla_\n + g_{\m\n} )\phi = 0\,.
\ee
There is a solution similar to the Schwarzschild-de Sitter black hole. In global coordinates, the metric and dilaton are
\be
ds^2 = \f{1}{\cos^2 \s} (-d\s^2 + d\varphi^2)\,,\qquad \phi = \phi_r \frac{\cos\varphi}{\cos\sigma} \ ,
\ee
with $\s \in \left(-\f{\pi}{2}, \f{\pi}{2}\right), \,\varphi \in (0,2\pi)$, and $\phi_r>0$. The Penrose diagram is given in figure \ref{fig:dsisland}. 

In patch 2, the dilaton approaches $+\infty$ as $\sigma \to \frac{\pi}{2}$, so this is the asymptotic future boundary ${\cal I}^+$. In patch 1, the dilaton approaches $-\infty$ as $\sigma \to \frac{\pi}{2}$, which is viewed as the black hole singularity. 
In the embedding into three-dimensional Minkowski spacetime $ds^2 = -dX_0^2 + dX_1^2 + dX_2^2$,   global coordinates are defined by
\be
X_0 = \tan \s\,,\qquad X_1 = \f{\sin \varphi}{\cos \s}\,,\qquad X_2 = \f{\cos \varphi}{\cos \s} \ .
\ee
The solution in patch 2 is
\be
ds^2 = \f{1}{\sinh^2 T}(-dT^2 + dX^2)\,,\qquad \phi = -\phi_r \coth T\,,\qquad X \in \mathbb{R}\,, T < 0\,, \phi_r > 0\,,
\ee
with embedding into three-dimensional Minkowski spacetime given as
\be\label{hypembed}
X_0 = -\f{\cosh X}{\sinh T}\,,\qquad X_1 = -\f{\sinh X}{\sinh  T}\,,\qquad X_2 = -\coth T\,.
\ee
Notice that the dilaton diverges to $+\infty$ near the future boundary. In the higher-dimensional perspective this indicates that the size of the transverse sphere is diverging. From the embedding coordinates we can see that moving into the other hyperbolic patch, $\varphi \rightarrow \varphi + \pi$, is the same as $\{T, X\} \rightarrow \{-T+i\pi, -X\}$.  In patch 1, the metric is the same, but the dilaton has the opposite sign,
\be
\phi = \phi_r \coth T \ .
\ee

The single-interval entanglement entropy in global coordinates is derived by writing the metric for Euclidean dS$_2$ as Weyl-equivalent to flat spacetime:
\be
ds^2 =\Omega^{-2}dx d\bar{x} \,,\qquad \Omega =  \f 1 2(1+ x \bar{x}) \,. 
\ee
This means we can insert the appropriate Weyl factors to transform the flat-spacetime answer and obtain
\be
\Ssemi = \f c 6 \log \left(\f{(x_1-x_2)(\bar{x}_1 - \bar{x}_2)}{\epsilonuv^2 \Omega(x_1)\Omega(x_2)}\right) 
\ee
for an interval with endpoints at $x_1$ and $x_2$, where $c$ is the central charge of the CFT.
The Lorentzian global coordinates are given as 
\be\label{xglobal}
x = e^{-i(\s - \varphi)}\,,\qquad \bar{x} = e^{-i(\s + \varphi)}
\ee
leading to 
\be
\Ssemi = \f c 6 \log \left(\f{2(\cos (\s_1-\s_2)-\cos(\varphi_1-\varphi_2))}{\epsilonuv^2 \cos(\s_1) \cos(\s_2)}\right)\,.
\ee
This is the entanglement entropy for the Hartle-Hawking state on dS$_2$. To get the entanglement entropy in the hyperbolic patch we use the coordinate transformations
\be\label{globalhyp}
\s = \tan^{-1}\left(-\f{\cosh X}{\sinh T}\right)\,,\qquad \varphi = \tan^{-1} \left(\f{\sinh X}{\cosh T}\right)\,.
\ee
With the standard branch of $\tan^{-1}$, this is the coordinate change for hyperbolic patch 2, and shifting $\varphi \to \varphi + \pi$ gives the coordinate change for patch 1. Our convention is such that each (future) hyperbolic patch is covered by $X \in \mathbb{R}$, $T<0$, with $T$ increasing to the future and $X$ increasing to the right. 
Using this coordinate transformation, we get the entanglement entropy for a single interval within one hyperbolic patch:
\be\label{entsingle}
\f c 6 \log \f{2(\cosh(X_2-X_1)-\cosh(T_2-T_1))}{\epsilonuv^2 \sinh T_1 \sinh T_2 }\,.
\ee
To move one of the points into the neighboring hyperbolic patch we use the continuation above to obtain
\be\label{hyper2}
\f c 6 \log \f{2(\cosh(X_2+X_1)+\cosh(T_2+T_1))}{\epsilonuv^2 \sinh T_1 \sinh T_2 }\,.
\ee
This is the entanglement entropy for one endpoint in patch 1, and the other endpoint in patch 2.

Although the setup seems different, this is actually very similar to our discussion of an FRW universe purified by an auxiliary spacetime in the thermofield double state, as in sections \ref{s:matterentropy}-\ref{s:recollapse}. The black hole interior, patch 1, plays the role of the FRW universe. This hyperbolic universe is in the thermofield double state with the exterior, so patch 2 is playing the role of the auxiliary spacetime in the previous discussion. The difference is that this region is now part of the physical spacetime, and has a dynamical dilaton and nontrivial conformal factor in the metric. The entanglement entropy \eqref{hyper2} is identical to the entanglement entropy of regions on opposite sides of the thermofield double in equation \eqref{sia1}, after inserting the appropriate conformal factors (up to a factor of 2, because \eqref{hyper2} only counts the entropy of one of the two intervals in $(I \cup R)^c$, while \eqref{sia1} counts both).

\subsection{Islands in dS$_2$}
With the above solutions we will exhibit an island in the crunching hyperbolic patch for a region in the neighboring hyperbolic patch. We pick the region $R$ to be in patch $2$ with
\be
R: \qquad \{T = T_R\,,\quad X \in [-X_R, X_R]\}\,,\qquad R \in \text{ patch 2}
\ee
and require $|T_R| \ll 1$ so that $R$ is anchored near ${\cal I}^+$. Since the Planck scale is diverging near $\mathcal{I}^+$ in this hyperbolic patch, we ignore the effects of gravity in region $R$, meaning that we will not include an area term from $\p R$ in the generalized entropy. 

Without any island contribution, region $R$ has von Neumann entropy
\be\label{ensemi}
\Ssemi(R) = \f c 3 \log \left(\f{2 \sinh X_R}{\epsilonuv \sinh (-T_R)}\right)  .
\ee
We consider the inclusion of an island region $I$ given as 
\be
I: \qquad \{T = T_I, \quad X\in[-X_I, X_I]\}\,,\qquad I \in \text{ patch 1}
\ee
Since the global state is pure, we can compute the generalized entropy by considering the two-interval region $(I \cup R)^c$, which gives
\be
S_{\rm gen}(I \cup R) = 2\phi_0 +2\phi_r \coth T_I + \f c 3 \log \left(\f{2(\cosh (X_R + X_I) + \cosh (T_R + T_I))}{\epsilonuv \epsilonrg \sinh T_I \sinh T_R}\right),
\ee
where we set $4G = 1$. The factor of two comes from the two intervals in $(I\cup R)^c$. We assume the entropy of these two intervals factorizes, which holds when the cross-ratio of the four points at the boundaries of $I$ and $R$ is in an appropriate OPE limit, as can be checked self-consistently at the end. We have also included the gravitational entropy $S_{grav} = \phi$ for each endpoint of the island region. This removes the UV divergence at the boundary of region $I$, as it has the effect of replacing $\epsilonuv \to \epsilon_{\rm rg}$ in the matter entropy where $\epsilonrg$ is an RG scale, as discussed in section \ref{ss:condition1}.  Note that we define the generalized entropy without any gravitational term at $\p R$, so it does not remove the UV divergence there. 

Extremizing the island region with respect to $X_I$ gives a minimum at 
\be
X_I = -X_R
\ee
 and extremizing with respect to $T_I$ gives a maximum, which as $T_R \to 0$ is given as
\be\label{islandtime}
T_I \approx -\sinh^{-1} \f{6\phi_r}{c} \ . 
\ee
At leading order in small $T_R$ the island entropy is 
\be
S_{\rm island}(R) = 2(\phi_0+ \phi_I)+\f c 6 \log\left(\f{4(\phi_I - c/6)}{T_R^2 \epsilonuv^2 \epsilonrg^2 (\phi_I+c/6)}\right)
\ee
where
\be
\phi_I = -\phi_r \sqrt{1+\left(\f{c}{6\phi_r}\right)^2}\,.
\ee
This entropy is independent of $X_R$, while the semiclassical entropy \eqref{ensemi} grows linearly at large $X_R$, so at some length the inclusion of the island will minimize the generalized entropy. The critical value of the length where the transition occurs -- which we will call the Page length in analogy to the black hole context -- can be found by solving $S_{\rm island}(R) = \Ssemi(R)$ and gives 
\be\label{xpage}
X_{\rm Page} = \sinh^{-1}\left(e^{\f 6 c (\phi_0+\phi_I)}\frac{1}{\epsilonrg}\sqrt{\f{\phi_I - c/6}{\phi_I + c/6}}\right) \ .
\ee
For large regions, this reduces to $X_{\rm Page} \approx 6(\phi_0+\phi_I)/c$.

Notice that the quantum state in region $R$ is not accessible to a single observer in de Sitter space. We can modify the problem so as to exit from the de Sitter epoch. We can do this locally in patch 2 by extending past $\mathcal{I}^+$ and gluing on a flat-spacetime hat. An observer that goes into the flat spacetime hat can have region $R$ in her past lightcone. This observer is sometimes called a ``census taker" \cite{Susskind:2007pv}. We will consider a closely related problem in the next section.

An important aspect of the solution we have considered above is that the dilaton diverges to $-\infty$ at $\mathcal{I}^+$ in patch 1 while it diverges to $+\infty$ at $\mathcal{I}^+$ in patch 2. This means that patch 2 is inflating toward the future while patch 1 is crunching. This is because this solution can be thought of as a dimensional reduction of Schwarzschild-de Sitter in an extremal limit, also known as the Nariai solution. Patch 1 is the black hole interior, and the island we have exhibited is very analogous to the islands exhibited in black hole solutions in other spacetimes. This feature played a key role in the calculation above, since the gravitational cost for nucleating an island decreased as we moved toward $\mathcal{I}^+$ in patch 1, whereas for an inflating patch the cost would increase.\footnote{Notice that from a higher dimensional perspective, $\phi_0$ is set by the cosmological constant, unlike the case for anti-de Sitter spacetime where it can be arbitrarily large.}
%\enlargethispage{0.5cm}%ad hoc spacing to keep footnote on page

\subsection{Bekenstein area bound and the quantum normal regions}\label{sec:bekqnorm}
We would like to check that the Bekenstein area bound is violated as discussed in section \ref{ss:condition1}. We consider an interval centered around the origin in the hyperbolic patch 1 with $X \in (-X_I, X_I)$ at time $T_I$. The regularized matter entropy is given from \eqref{entsingle} as
\be
\Sred(I) = \f c 3 \log \f{2 \sinh X_I}{\epsilonrg \sinh(- T_I)  }\,,
\ee
while the gravitational entropy is given by
\be
\f{\text{Area}(\partial I)}{4} = 2(\phi_0 + \phi_r \coth T_I)\,.
\ee
The Bekenstein area bound is violated when
\be
-\f{\sinh X_I}{\sinh T_I} \gtrsim \epsilonrg \exp\left(\f 6 c (\phi_0+\phi_r \coth T_I)\right).
\ee
For large $X_I$, with $T_I$ given by \eqref{islandtime}, this inequality becomes
\be
X_I \gtrsim \f 6 c (\phi_0 + \phi_I) \,,
\ee
where we have ignored subleading logarithmic pieces. Up to these subleading pieces this is  equal to the Page length \eqref{xpage}.

\begin{figure}
\begin{center}
\includegraphics{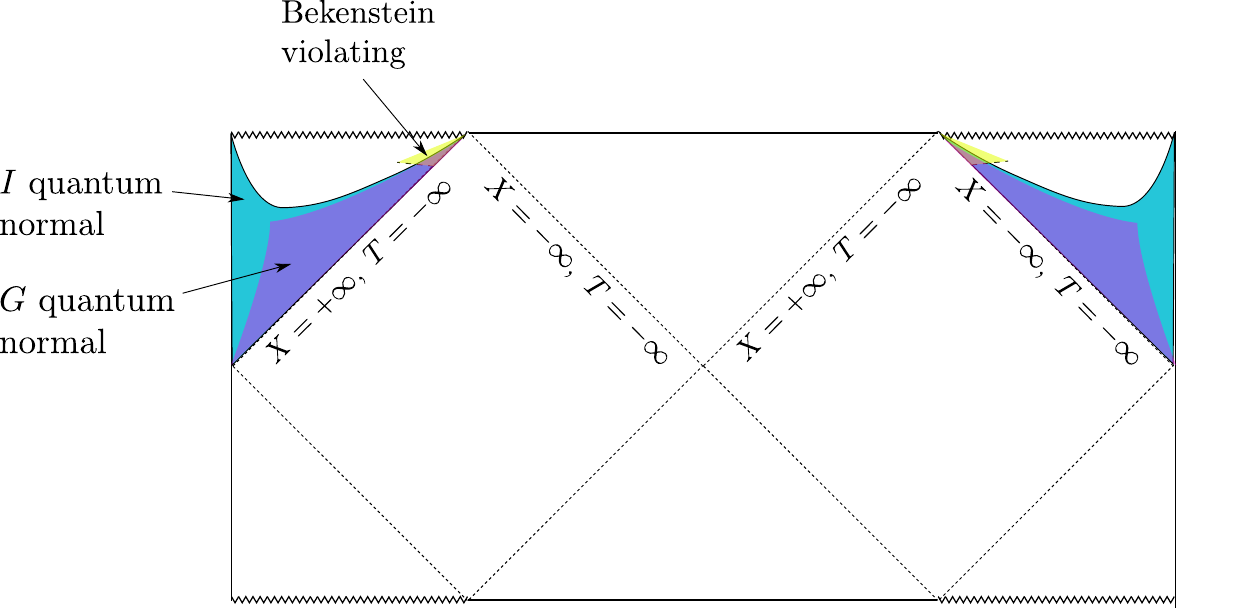}
\end{center}
\caption{Regions of the dS$_2$ spacetime, assuming a reflection-symmetric region $I$ within patch 1. The endpoints of $I$ must fall in the overlaps of the quantum normal region and Bekenstein-violating region. Only the semiclassical portion of the Bekenstein-violating region is drawn, which is why it appears to end abruptly.\label{fig:dSregions}}
\end{figure}

The endpoints of our island must be in the quantum normal regions of $I$ and $G$, as discussed in section \ref{sec:qnormal}-\ref{sec:qnormal2}. We first consider the quantum normal region for $I$. We consider the future-directed outgoing null derivative of the right endpoint $X_I$. To take this derivative we will need the formula for the entanglement entropy of a single interval in the hyperbolic patch, given in \eqref{entsingle}. Altogether the generalized entropy is given as 
\be
S_{\rm gen}(I) = 2 \phi_0 + \phi_r \coth T_1 + \phi_r \coth T_2 + \f c 6 \log \f{2(\cosh (X_2-X_1) - \cosh(T_2 - T_1))}{\epsilonrg^2 \sinh T_1 \sinh T_2}\,.
\ee
 Differentiating $S_{\rm gen}(I)$ with respect to $X_2^{\pm} = T_2 \pm  X_2$ and setting $X_2 = - X_1 = X_I$ and $T_2 = T_1 = T_I<0$ gives the pair of inequalities
\be
\coth X_I \geq \pm \left( \coth T_I + \f{6 \phi_r}{c \sinh^2 T_I}\right)
\ee
where we have required the outgoing derivative to be non-negative and the ingoing one to be non-positive. Since $X_I>0$ and therefore $\coth$ is monotonically decreasing we have
\be\label{qnrds}
X_I \leq \coth^{-1}\Big| \coth T_I + \f{6\phi_r}{c \sinh^2 T_I}\Big|.
\ee
This is the quantum normal condition for region $I$. Notice that since our island region is at arbitrarily large $X_I = -X_R $, we need this bound to trivialize, i.e. the right hand side needs to diverge for the existence of the island to be consistent. That happens when the argument of $\coth^{-1}$ is less than one, in which case there is no upper bound. This restricts
\be\label{trest}
\log \sqrt{1-12\phi_r/c} < T_I < \log \sqrt{\f{1}{1+12\phi_r/c}}\,.
\ee
The first inequality is trivial for $\phi_r/c>1/12$.  Our island does indeed lie in this range. Interestingly, this becomes very restrictive for $\phi_r/c \ll 1$, forcing $T_I \approx -6\phi_r/c$. In the regime $\phi_r/c\ll 1$ this is precisely what we found for the island time, $T_I = -\sinh^{-1} (6\phi_r/c) \approx -6\phi_r/c$. It is also restrictive as $\phi_r/c \gg 1$, requiring $T_I \lesssim \log \sqrt{c/(12\phi_r)}\rightarrow -\infty$.

Now we consider the quantum normal condition for the region $G$ defined as the interior patch minus the island. That is,
\be\label{gdef}
G =   [-\infty, -X_I] \cup [X_I, \infty] \ ,
\ee
at arbitrary time $T_I$, in patch 1. This is similar to our choice of region $G$ in the FRW thermofield double, where it was the gravitating region minus the island. We will assume the entanglement entropy factorizes into twice the entanglement entropy of one of the two intervals, which is a good approximation for $X_I \gtrsim 1$.\footnote{The OPE limit we are taking is when the size of the two intervals comprising $G$ is small compared to their separation. The cross-ratio can be computed using the Minkowski distances in the embedding space, since this is the same as the distance in the metric $dxd\bar{x}$ with $x$, $\bar{x}$ given by \eqref{xglobal}. Using the embedding \eqref{hypembed}, the cross-ratio simplifies to $\exp(-2X_I)$, which is small for $X_I \gtrsim 1$.   
} 
\enlargethispage{0.5cm}%ad hoc spacing to keep footnote on page
The formula for the generalized entropy is therefore 
\be\label{hrrg}
S_{\rm gen}(G) = 2\phi_r \coth T_I - \frac{c}{3} X_I - \frac{c}{3}\log \sinh T_I + \mbox{const.}
\ee
The quantum normal condition is
\be
(\mp \p_{T_I} - \p_{X_I})S_{\rm gen}(G) \geq 0 
\ee
which simplifies to
\be\label{qnrds2}
\Big| \coth T_I + \f{6\phi_r}{c \sinh^2 T_I}\Big| \leq 1 \ .
\ee
This is the same range we found below \eqref{qnrds} by requiring that there is no quantum apparent horizon for region $I$, so in fact both of the quantum normal conditions (for $I$ and $G$) are satisfied in this range.

\begin{figure}
\begin{center}
\includegraphics[scale = 2]{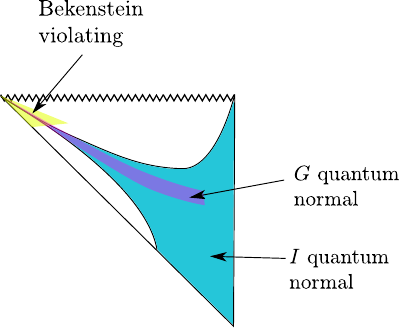}
\end{center}
\caption{Half of the hyperbolic patch of dS$_2$, with the various regions displayed for $\phi_r/c\ll 1$. In this limit, the range of allowed times for the endpoint of an island becomes very narrow.\label{fig:dSsingleregion}}
\end{figure}

The quantum normal region (assuming $12\phi_r/c > 1$) is shown in figure \ref{fig:dSregions} together with the Bekenstein-violating region. For $12 \phi_r/c < 1$ there is an additional excluded region coming out of the right horizon $X=T=-\infty$ and ending at the corner $X=-\infty$, $T = $ finite. In the limit $\phi_r/c \ll 1$ this is shown in figure \ref{fig:dSsingleregion}.

\subsection{Pure de Sitter in higher dimensions?}

Our dS$_2$ model is similar to a black hole in a higher-dimensional de Sitter spacetime. What about pure de Sitter in higher dimensions? Here we can  make progress by analyzing the Bekenstein area bound, our condition 1. In the Hartle-Hawking state, the matter entropy is computed by a Weyl transformation from flat spacetime. This means the matter entropy does not scale with volume -- it grows with $r_I$ at the same rate as the gravitational entropy. Therefore the Bekenstein area bound will not be violated, so no islands appear in pure de Sitter spacetime.

\section{Minkowski bubble in dS$_2$}\label{sec:JTglue}

One physical motivation for considering crunching cosmologies is in the context of eternal inflation. For any quantum-gravitational theory with a landscape, like string theory, gravitational instantons lead to the nucleation of bubbles in a parent inflating cosmology. This process may continue ad infinitum, stopping locally only when ``terminal vacua" are reached. These are  bubbles with vanishing or negative cosmological constant. The bubbles with negative cosmological constant generically lead to a crunch. 
In the previous sections we focused on the presence of islands in the context of the thermofield double, where one of the two spacetimes is a crunching cosmology. The island region of the crunching cosmology was encoded in a region in the second, Minkowski spacetime. Here, we would like to consider a similar setup where the Minkowski spacetime is the terminal vacuum in which the observer lives and the crunching cosmology is a neighboring bubble. The necessary conditions outlined for the presence of an island in section \ref{s:generalconditions} are independent of region $R$, so it is at least plausible that islands may exist in this setup as well. 

One difference with the previous thermofield double examples has to do with the predominant decay channel considered in the literature, the Coleman-De Luccia instanton \cite{Coleman:1980aw}. This solution preserves $O(D)$ symmetry and is expected to be the dominant decay channel, i.e. it is the bounce of lowest Euclidean action. The $O(D)$ symmetry means that the spatial slices are hyperboloids and we need to consider an open FRW universe, while in sections \ref{s:matterentropy}-\ref{s:recollapse} we have been considering flat spatial slices. 

It is difficult to compute the entropies needed to look for islands explicitly in a higher-dimensional model of bubble nucleation. So we instead consider a two-dimensional model. Rather than compute an instanton that mediates decay and solve for the global geometry, we use Jackiw-Teitelboim gravity in dS$_2$ plus a large-$c$ matter CFT. We will glue this solution to flat spacetime to mimic a bubble nucleation of vanishing cosmological constant, so we will take the Minkowski theory to be flat-spacetime JT gravity. (Similar calculations of islands in the context of dS$_2$ glued to flat spacetime have been presented in \cite{cgmTalks}.) This sharp gluing mimics vanishingly thin bubble walls. The two theories and their solutions in the relevant patches are 
\begin{align}
S = \f{1}{16\pi G}\int d^2 \sqrt{-g} \,\phi (R-2)+S_{CFT}\,,\qquad ds_{dS}^2 = d\theta^2 - \sin^2 \theta dt^2\,,\qquad \phi_{dS} = \phi_m \cos \theta\\
S =\f{1}{16\pi G}\int d^2 \sqrt{-g}\,( \phi R -2)+S_{CFT}\,,\qquad ds_{flat}^2 = dr^2 - r^2 dt^2 \,,\qquad \phi_{flat} =  \phi_0 + \f{ r^2}{2}
\end{align}
Notice that the sign of the cosmological constant in the flat-spacetime JT action is different than the usual one \cite{Callan:1992rs}. As we will see in \eqref{hypflat} this is so that the dilaton grows toward $\mathcal{I}^+$. 

\begin{figure}[t]
\begin{center}
\includegraphics[scale=.4]{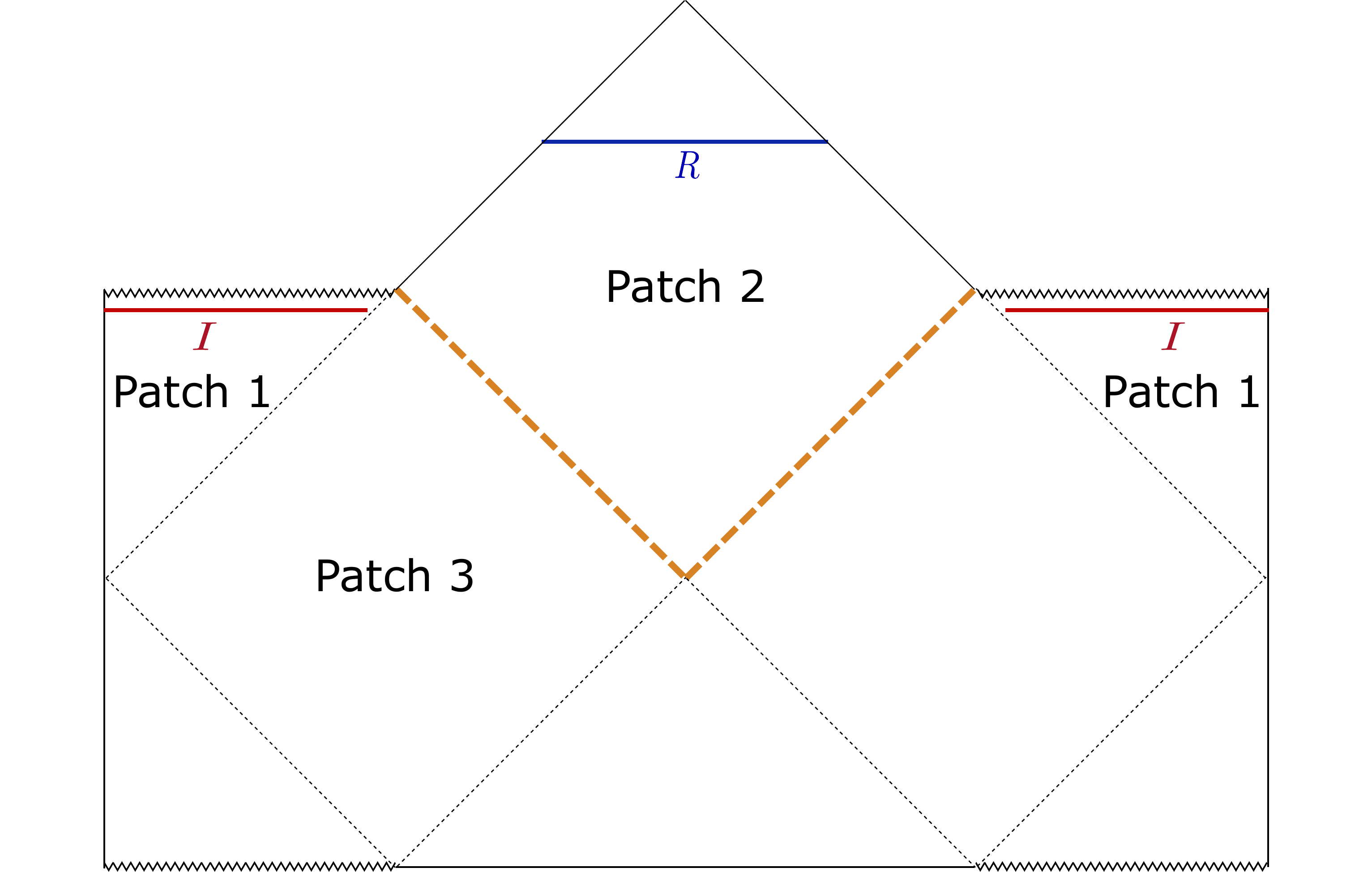}
\end{center}
\caption{The Penrose diagram of dS$_2$ glued to flat spacetime along the thick dashed orange line. Patch 1 is a hyperbolic patch of dS$_2$, patch 2 is a hyperbolic patch of flat spacetime, and patch 3 is a static patch. We calculate the von Neumann entropy of region $R$ and find the island $I$.} \label{fig:multiverse}
\end{figure}

The walls of a Coleman-De Luccia bubble are timelike and accelerate outward to quickly become nearly null. To simply model this we will take a null limit and glue across $\theta = r = 0$, represented by the thick dashed orange line in figure \ref{fig:multiverse}, which means we need the continuation of the flat spacetime solution into the hyperbolic patch: 
\be\label{hypflat}
ds^2_{flat} = -dt^2 + t^2 dX^2 = e^{2T}(-dT^2 + dX^2)\,,\qquad \phi_{flat} = \phi_0+\f{ t^2}{2} = \phi_0 +\f{ e^{2T}}{2} \,.
\ee
The gluing requires picking $\phi_0 = \phi_m$. As discussed in the previous section, the neighboring de Sitter region is considered to be a crunching cosmology since the dilaton is diverging to $-\infty$ in the future.

We would like to compute the entropy of a region $R$ in patch 2, with and without the inclusion of an island. We take region $R$ to be at a fixed time $T_R$ and with spatial extent $(-X_R, X_R)$. 
 We can compute the semiclassical entropy of the island by using the coordinate transformation 
\be
t = e^T \cosh X\,,\qquad x = e^T \sinh X
\ee
to transform the Minkowksi vacuum answer into
\be\label{eehyperbolic}
\Ssemi = \f c 6 \log \left(\f{(\Delta x)^2-(\Delta t)^2}{\epsilonuv^2}\right) = \f c 3 \log\left(\f{2 \sinh X_R}{e^{-T_R}\epsilonuv}\right) \ .
\ee 
Now we would like to consider the inclusion of an island. Since this problem is structurally very similar to the one in the previous section, we again look for an island in patch 1. We assume the two-interval entanglement entropy on $(I \cup R)^c$ factorizes. The entropy for one of the two intervals in $(I \cup R)^c$ is given by the same formula as \eqref{hyper2}, except we need to take into account the different Weyl factor in patch 2. The gravitational entropy remains the same since patch 1 is unchanged. Thus the generalized entropy on $I \cup R$  is  given by
\be
S_{\rm gen} = \f c 3 \log\left(\f{2(\cosh (X_R+X_I)+\cosh(T_R+T_I))}{- e^{-T_R}\sinh T_I\,\epsilonuv \epsilonrg}\right)+2(\phi_0 + \phi_r \coth T_I),
\ee
where we set $4G = 1$. Extremizing this with respect to $X_I$ gives $X_I = -X_R$. Extremizing with respect to  $T_I$ gives
\be
\coth T_I - \tanh\f{T_I+T_R}{2} + \f{6\phi_r}{c \sinh^2 T_I} = 0\,.
\ee
Since we want the endpoints of region $R$ to sit on $\mathcal{I}^+$ we need to take $T_R \rightarrow \infty$, which leads to the solution
\be
T_I \approx -\f 1 2 \log \left(1+\f{12\phi_r}{c}\right) \  .
\ee
 Notice that this saturates the upper bound \eqref{trest} and is therefore on the border of the region allowed by quantum normalcy of $G$. The island entropy is given by 
\be
S_{\rm island}(R) \approx 2(\phi_0 + \phi_I) +\f{2 c T_R}{3}-\f{c}{3}\log \epsilon_{\rm uv}
\ee
where
\be
\phi_I = -\phi_r\left(1+\f{c}{6\phi_r}\right).
\ee
Thus as long as $\phi_0 \gg \phi_r(1+c /\phi_r)$  we can remain weakly coupled. The critical value of the length where a transition between the semiclassical entropy and the island entropy occurs can be found by solving $S_{\rm island}(R) = \Ssemi(R)$, which gives 
\be\label{xpage2}
X_{\rm Page} \approx \sinh^{-1}\left(e^{\f 6 c (\phi_0+\phi_I)+T_R}\right)
\ee
Notice that since patch 1 is unchanged from our discussion in section \ref{s:dS2}, we satisfy the Bekenstein area bound and the endpoints of our island will be in the quantum normal region, as in section \ref{sec:bekqnorm}. For $\phi_r/c \ll 1 $ we find $T_I \approx -6\phi_r/c$, as dictated by the shrinking of the quantum normal region in this limit. 

To make this case closer to the one in the previous section, we can regulate region $R$ to sit near but not on $\mathcal{I}^+$. Then as we increase the length of region $R$ we will reach the Page length where the island solution dominates. In this case, the quantum state on region $R$ is in the past lightcone of an observer that goes to future timelike infinity of patch 2. The unbounded growth of entropy with respect to increasing the size of region $R$, which in the previous section was only accessible to a metaobserver with access to spacelike patches of $\mathcal{I}^+$, is now accessible to a single observer (the census taker \cite{Susskind:2007pv}). Conceptually, a Minkowski hat provides an ``exterior" view of cosmology similar to the exterior view of a black hole. 

This spacetime has a maximal analytic extension with an infinite number of inflating regions and crunching regions, just like Schwarzschild-de Sitter in higher dimensions. It is tempting to use this extension as a better model for a multiverse.

\section{Tensor network picture}\label{s:tensors}
We will now discuss a simple toy model for the island rule using tensor networks. This provides some intuition for the structure of the quantum state that leads to an island.  Similar toy models were discussed in \cite{1910.00972,1912.00909, 1807.06041, 1912.00228, 2006.12601, 2003.03406,2007.02987}.\footnote{See \cite{2002.05734} for a more detailed proposal for the structure of the quantum state responsible for islands. }
The tensor network does a nice job of capturing the intuition for deformations of $I$ in spacelike directions, and for the violation of the Bekenstein area bound, but it does not do so well in the time direction. There is no apparent analogue of the extremality condition or the quantum normal condition. 
The purpose of this section is to provide some basic intuition for islands and the Bekenstein bound.

\subsection{Evaporating black hole}
\begin{figure}
\begin{center}
\includegraphics[scale=1]{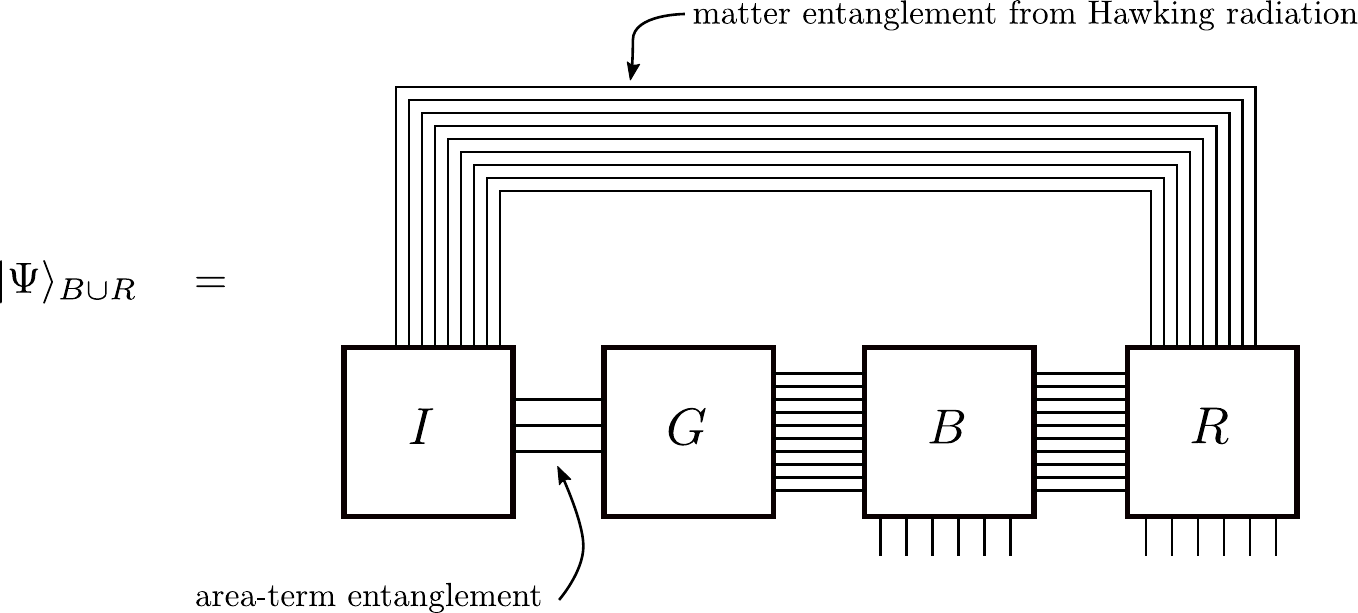}
\end{center}
\caption{Tensor network toy model for the quantum state of an evaporating black hole at late times. \label{fig:blackholetensors}}
\end{figure}

A tensor network that qualitatively captures the structure of the quantum state of an evaporating black hole at late times is shown in figure \ref{fig:blackholetensors}. The regions $I,G,B,R$ roughly match the corresponding regions in the black hole in figure \ref{fig:evap-regions}.

Each box represents a tensor; bonds joining the tensors are contracted indices, and free legs are uncontracted indices. The tensor network is a quantum state in the Hilbert space associated to the uncontracted indices.\footnote{So the diagram represents the quantum state 
\be
\sum 
I_{j_1j_2\cdots}^{k_1k_2\cdots}
G_{k_1k_2\cdots}^{\ell_1\ell_2\cdots}
B^{i_1i_2\cdots}_{\ell_1\ell_2\cdots, \sigma_1\sigma_2\cdots} R^{j_1j_2\cdots}_{i_1i_2\cdots, \alpha_1\alpha_2\cdots}|\sigma_1\rangle|\sigma_2\rangle\cdots |\alpha_1\rangle|\alpha_2\rangle\cdots \ .
\ee
} Note that there are no external legs on $I$ or $G$ because these represent the gravitational regions. For the purposes of the toy model we assume gravity is unimportant in regions $B$ and $R$ and treat them like a quantum field theory.

The short bonds correspond to the spatial geometry of a late-time slice in figure \ref{fig:evap-regions}. The long bonds connecting $I \leftrightarrow R$ correspond to matter entanglement, \ie,  the long-range entanglement between Hawking radiation in region $R$ and its interior partners in region $I$.

For generic tensors, the von Neumann entropy $S(\rho_R)$ in this quantum state is proportional to the length of the minimal cut that separates $R$ from the rest of the diagram. This feature  is reminiscent of the Ryu-Takayanagi formula \cite{hep-th/0603001} and is the starting point for an intriguing correspondence between holography and tensor networks \cite{0905.1317,1811.05382,1811.05171}. Although the tensor network is discrete and non-dynamical it has an uncanny ability to predict complicated gravitational phenomena, including quantum extremal islands.

At early times, the area of the black hole is large, while the entanglement between the interior and the Hawking radiation is small. Therefore we can estimate the entropy by a minimal cut that simply excises region $R$:
\be
\includegraphics{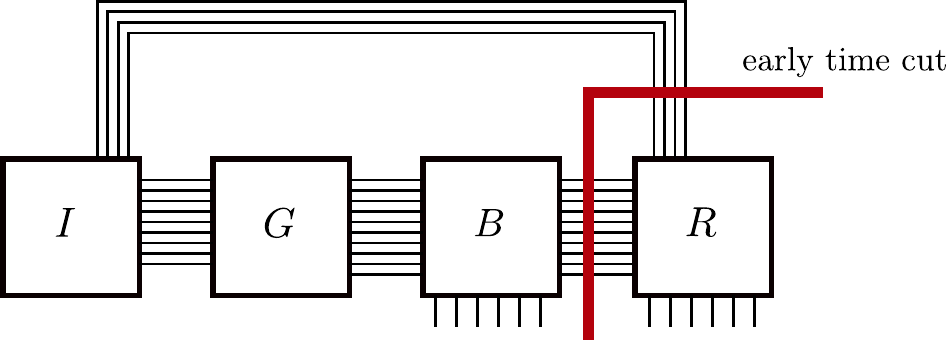}
\ee
The entropy from this cut grows with time as more Hawking radiation enters region $R$. Meanwhile the area term, represented by the links $I \leftrightarrow G$, shrinks as the black hole evaporates. Eventually, it is more economical to cut along $\partial I$:
\be
\includegraphics{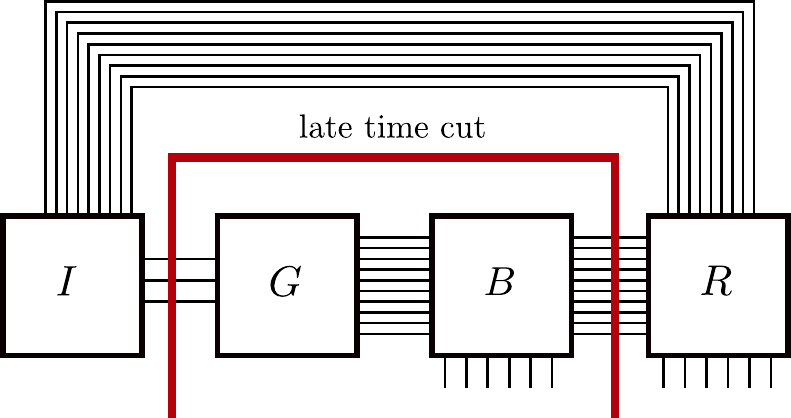}
\ee
The entropy from this cut decreases with time, because the contribution from the bonds $I \leftrightarrow G$ is proportional to the shrinking black hole area.

The transition from one cut to another occurs at the Page time and indicates the formation of a quantum extremal island. The toy model illustrates one way for an island $I$ to be `encoded' in an auxiliary system $R$. A more detailed understanding of this encoding can be found in the language of quantum error correction \cite{1411.7041, 1503.06237, 1607.03901, 1911.11977, 2006.08002}.

\begin{figure}[h]
\begin{center}
\includegraphics[scale=1]{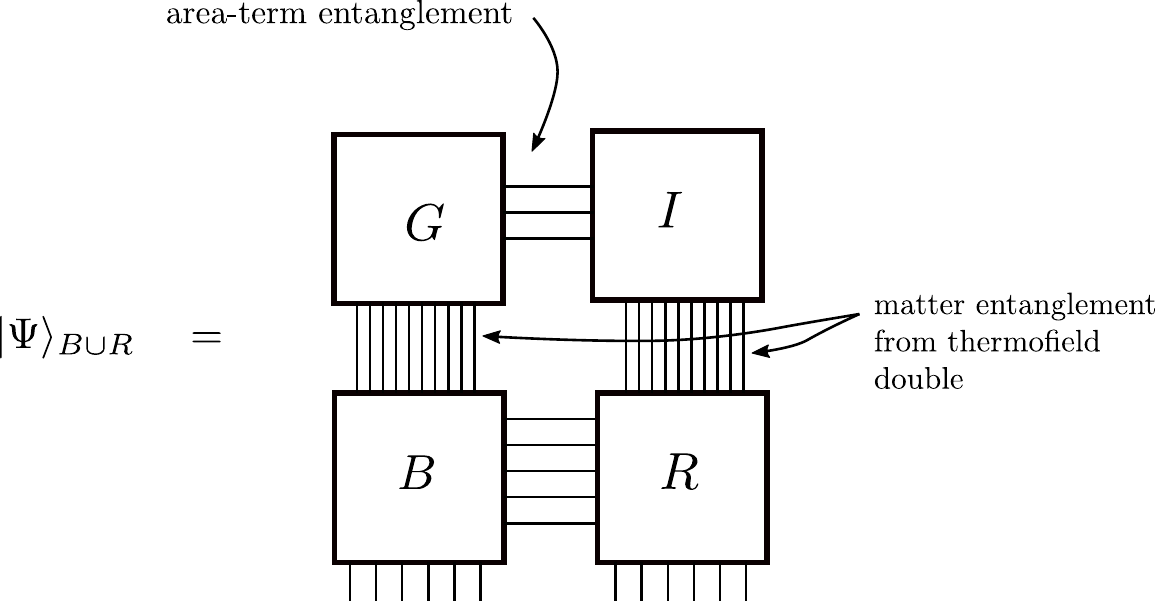}
\end{center}
\caption{\small Tensor network toy model for the thermofield double of FRW. \label{fig:frwtensors}}
\end{figure}

\subsection{Tensor network for FRW}

A tensor network toy model for FRW is in figure \ref{fig:frwtensors}. It is identical to the black hole, but the interpretation of the bonds is different, and we have reorganized the picture to match the geometry of the FRW+Minkowski thermofield double. The region labels match those in figure \ref{fig:frwisland}. Now the entanglement between $I$ and $R$ comes from our choice of the thermofield double state, rather than from Hawking radiation. This entanglement is time-independent for comoving regions, while the area-term entanglement depends on time through the scale factor.

This picture highlights the similarities between the evaporating black hole and FRW. The model also clearly has an analogue of our condition (1) in the introduction, \ie, islands must violate the Bekenstein area bound -- the matter entropy of $I$ counts vertical tensor legs on $I$, and the gravitational entropy counts horizon tensor legs on $I$. In the toy model there is no need to worry about regulating the quantum area. 

On the other hand, from the tensor picture we might expect to find islands in FRW with radiation only. This is not the case, a result that we traced back to the fact that in radiation-only FRW, the Bekenstein-violating region has no overlap with the quantum normal region. Apparently the tensor network picture for FRW succeeds in predicting the island only near the turning point of a recollapsing cosmology.

%\enlargethispage{0.5cm}%ad hoc spacing to avoid orphan
Earlier network-like toy models for cosmology which also include dynamics can be found in \cite{quant-ph/0501135,1702.06959}.

\ \\
\bigskip

\noindent \textbf{Acknowledgments} We thank Ahmed Almheiri, Kanato Goto, Raghu Mahajan, Juan Maldacena, Liam McAllister, Mudassir Moosa, and Amir Tajdini for helpful discussions. The work of YJ and ES is supported by the Simons Foundation through the Simons Collaboration on the Nonperturbative Bootstrap. The work of TH is supported by DOE grant DE-SC0020397.

\appendix

\section{Entanglement across the thermofield double in 2d CFT}\label{app:cft2tfd}

In this appendix we consider some properties of the matter entanglement in a 2d CFT, for two regions on opposite sides of the thermofield double. This setup is the same as section \ref{s:matterentropy} but in two dimensions we can be more explicit.

Consider a 2d CFT in two copies of Minkowski spacetime, in the thermofield double state. Define a region $I$ in system 1, and region $R$ in system 2:
\begin{align}
I: &\{ x_1 \in [-r_I, r_I], t_1=t_I \} \\
R: &\{ x_2 \in [-r_R, r_R], t_2= t_R \} \notag
\end{align}
In this subsection the goal is to compute the CFT entropy of $I\cup R$.
We can think of this theory as living on an analytic continuation of the Euclidean cylinder. Take the complex cylinder coordinate $z$ with $z \sim z + i \beta$. Then the two systems live at Im $z = \pm i \beta/4$. The map to the plane is
\be
w = e^{2\pi z / \beta} \ .
\ee
The endpoints of region $I$ are $[z_1, z_2]$ and the endpoints of region $R$ are $[z_3, z_4]$, with
\begin{align}
z_1 &= -r_I-t_I + i \beta/4  & \bz_1 &= -r_I + t_I - i\beta/4 \\
z_2 &= r_I - t_I  + i \beta/4 & \bz_2 &= r_I +t_I - i \beta/4 \notag\\
z_3 &= -r_R-t_R - i \beta/4  & \bz_3 &= -r_R + t_R + i\beta/4 \notag\\
z_4 &= r_R - t_R  - i \beta/4 & \bz_4 &= r_R +t_R + i \beta/4  \notag
\end{align}
We will calculate the Renyi partition function $Z_n = \tr (\rho_{I \cup R})^n$ using the twist operator methods of Cardy and Calabrese \cite{Calabrese:2004eu, 0905.4013}, as applied to this problem in \cite{1303.1080} (see also \cite{0905.2069,1011.5482,1303.6955}). The only difference compared to \cite{1303.1080} is that here we allow regions $R$ and $I$ to be different sizes. The partition function is 
\begin{align}
Z_n &= \langle \sigma(z_1) \bsigma(z_2) \bsigma(z_3) \sigma(z_4) \rangle_{cyl} \\
&= \left( 2\pi \over \beta \right)^{8h_n} |w_1 w_2 w_3 w_4|^{2h_n}
\langle \sigma(w_1) \bsigma(w_2) \bsigma(w_3) \sigma(w_4) \rangle_{plane} \notag 
\end{align}
where $\sigma$ and $\bsigma$ are twist operators, with chiral scaling dimension
\be
h_n = \frac{c}{24}(n-1/n) \ .
\ee
The calculation of the twist correlator depends on the CFT.  For concreteness we will assume the CFT is holographic (i.e. has large central charge $c \gg 1$ and a large spectral gap), but the results are independent of this assumption in the kinematic regime we are ultimately interested in.  In a holographic CFT, the twist correlator is the minimum of two factorized channels. In the first channel we contract the points across the thermofield double and find
\begin{align}
Z_n &= \left( 2\pi \over \beta \right)^{8h_n} |w_1 w_2 w_3 w_4|^{2h_n} 
|w_1- w_3|^{-4h_n} |w_2 - w_4|^{-4h_n} \ . 
\end{align}
Plugging in the kinematics above and taking $n \to 1$ to compute the von Neumann entropy we find
\be\label{sia1}
S_1(I \cup R) 
= \frac{c}{3}\log \left[ \frac{\beta^2}{2\pi^2 \epsilonuv^2} \left(
\cosh \frac{2\pi(t_R - t_I)}{\beta} + \cosh \frac{2\pi(r_R - r_I)}{\beta} 
\right) \right]  \ .
\ee
The other channel is where we contract each twist operator with its partner on the same side of the TFD. In this channel the partition function is
\be
Z_n = \left( 2\pi \over \beta\right)^{8h_n}|w_1 w_2 w_3 w_4|^{2h_n}|w_1-w_2|^{-4h_n}|w_3-w_4|^{-4h_n}\ , 
\ee
and this leads to an entropy which is simply the sum of two thermal entropies,
\be\label{sia2}
S_2(I \cup R) = \frac{c}{3}\log \left[ \frac{\beta}{\pi \epsilonuv}\sinh \frac{2\pi r_R}{\beta} \right]
+\frac{c}{3}\log \left[ \frac{\beta}{\pi \epsilonuv}\sinh \frac{2\pi r_I}{\beta} \right] \ .
\ee
The full answer for a holographic CFT is the minimum of \eqref{sia1} and \eqref{sia2}. Except very near the transition where \eqref{sia1} and \eqref{sia2} are equal, this result actually applies to any 2d CFT, because we have just used to the OPE of the twist operators to approximate the partition function.

\section{Derivation of the timelike-maximum requirement}\label{app:maximin}
In this appendix we will show that the entropy $S_{\rm gen}(I \cup R)$ increases (or rather does not decrease) at second order under any timelike deformation of $\p I$, by taking two derivatives of the generalized entropy. We assume the quantum focusing conjecture (QFC) \cite{1506.02669} and entanglement wedge nesting (EWN) \cite{1204.1330, 1211.3494, 1610.08968}.  These are not really independent assumptions \cite{1610.08968} but we will use them both as inputs.  As discussed in section \ref{ss:maximin}, the same conclusion follows from the quantum maximin prescription derived  in \cite{1912.02799}.

Consider an island in 2d. Before extremizing, the generalized entropy is a function 
\be
S_{\rm gen}(I\cup R) = S(i^+, i^-, r^+, r^-)
\ee
where $(i^+, i^-)$ is an endpoint of $I$ and $(r^+,r^-)$ is an endpoint of $R$. It can also depend on other endpoints of $I$ and $R$ but this dependence is suppressed in the notation as any other endpoints are held fixed. The extremality condition is
\be
\p_{i^{\pm}} S(i^+, i^-, r^+, r^-) = 0 \ .
\ee
Act on these two equations with the total derivatives $d/dr^\pm$ to find
\begin{align}\label{sfour}
\p_{r^+} \p_{i^+} S + \frac{\p i^+}{\p r^+} \p_{i^+}^2 S  + \frac{\p i^-}{\p r^+} \p_{i^-} \p_{i^+} S&= 0
\end{align}
and three other similar equations with different combinations of derivatives.
QFC is the statement 
\be
\p_{i^+}^2 S \leq 0 \ , \qquad \p_{i^-}^2 S \leq 0 \ .
\ee
(Generally these would be covariant derivatives but here this is not necessary due to the extremality conditions.)
Pick orientations so that increasing $r^-$ grows $R$, increasing $r^+$ shrinks $R$. The orientations for the island are opposite, so increasing $i^-$ shrinks $I$ and increasing $i^+$ grows $I$. EWN requires that if we grow $R$, the island endpoint must move in a spacelike direction that grows $I$. Therefore
\be
\frac{\p i^+}{\p r^-} > 0 , \quad \frac{\p i^-}{\p r^-} < 0  \ .
\ee
And if we shrink $R$, the opposite holds, so
\be
\frac{\p i^+}{\p r^+} < 0 , \quad \frac{\p i^-}{\p r^+} > 0  \ .
\ee
The first term in \eqref{sfour} reduces to just the matter entropy contribution, because $\p_{r^{\pm}}\Area(\p I) = 0$. Thus strong subadditivity of the matter entropy implies
\begin{align}
\p_{i^+} \p_{r^-} S   < 0 \ , \quad \p_{i^-} \p_{r^+} S < 0  \ , \quad
\p_{i^+} \p_{r^+} S >0 \ , \quad \p_{i^-} \p_{r^-} S >0 \ .
\end{align}
Using SSA, EWN, and QFC (which can all be viewed as different aspects of QFC),
we see that the equation \eqref{sfour} takes the form
\begin{align}
(positive) + (positive) + (positive) \p_{i^-}\p_{i^+}S  = 0 \ .
\end{align}
Therefore
\be\label{pmbound}
 \p_{i^-}\p_{i^+}S \leq 0 \ .
\ee
The other three equations similar to \eqref{sfour} give the same sign constraint. In the notation of section \ref{ss:maximin}, this implies $\gamma \leq 0$ and therefore following the same steps as the classical argument we conclude that $S$ is non-decreasing at second order under timelike deformations.

The extension of this argument to higher dimensions is straightforward. We simply replace derivatives with respect to the endpoints by derivatives with respect to affine parameters $\lambda^{\pm}$ that deform the surface in null directions, along a small portion of the boundary with area $\mathcal{A}$. The quantum expansion in higher dimensions is defined by \cite{1506.02669}
\be
\Theta_{\pm} =\lim_{\mathcal{A} \to 0} \frac{4 }{\mathcal{A}} \frac{d S_{\rm gen}}{d\lambda^\pm} \ ,
\ee
which is a finite quantity.
The diagonal part of the quantum focusing conjecture requires 
\be
0 \geq \frac{d}{d\lambda^\pm} \Theta_{\pm}=\lim_{\mathcal{A} \to 0}\left[ \frac{d}{d\lambda^{\pm}}\left( \frac{4 }{\mathcal{A}} \right) \frac{d S_{\rm gen}}{d\lambda^\pm}+\frac{4 }{\mathcal{A}} \frac{d ^2 S_{\rm gen}}{d{\lambda^\pm}^2} \right] \ .
\ee
At a quantum extremal surface, the first term drops out due to the extremality condition $\frac{d S_{\rm gen}}{d\lambda^\pm}=0$. Therefore the argument goes through as above.

\section{Details of the CGHS/RST example}\label{app:cghs}
In this appendix we check the general conditions for the CGHS/RST model, as discussed in section \ref{ss:examples}. We will use the conventions of \cite{Fiola:1994ir, Hartman:2020swn}. The first example is the eternal black hole. Region $R$ is considered to be two equal intervals on $\mathcal{I}_L^+$ and $\mathcal{I}_R^+$ as drawn in figure \ref{fig:eternalBH}. For sufficiently late time an island was found to appear that stretched from the left horizon to the right horizon. The metric is given by $ds^2 = -e^{2\rho} dx^+ dx^-$ for some $\rho$ and $x^{\pm} = t \pm x$. The matter is in the $x^\pm$ vacuum, i.e. $T_{x^\pm x^\pm} = 0$. The entropy for a symmetric interval around the origin is therefore given by
\be\label{rstent}
S_{\rm mat} = \f c 3 \log \f{(x_2^+-x_2^-)}{\epsilon_{\rm uv} e^{-\rho(x_2)}}\,,
\ee
 where $(x_2^+, x_2^-)$ represents the right endpoint and $(x_2^-, x_2^+)$ represents the left endpoint. In this model, the combination $\rho + 6 S_{\rm grav}/c = \Omega +  k$, where $\Omega$ is a scalar field which characterizes the gravitational solution in a particular gauge and $k$ is some constant which can be ignored for our purposes. Using this equality and $\Omega = -x^+ x^-+M$ for the eternal black hole, the Bekenstein area bound becomes
\be
x_2^+ - x_2^- \gtrsim \exp\left(\f{6}{c}(M- x_2^+x_2^-)\right).
\ee
Working near the horizon $x_2^- = 0$ we see that for late enough time we can take the length $x_2^+ - x_2^-$ to be arbitrarily large and the inequality is obeyed.

We  now compute the quantum normal region for $I$.  This region will be a symmetric interval around the origin, and we will extremize with respect to the right endpoint. The quantum apparent horizon on the left will follow by symmetry. 
We will restrict to $t>0$ since that is where the island lives. The entanglement entropy in the eternal black hole background is given by \eqref{rstent}. Using $\rho + 6 S_{\rm grav}/c = \Omega+  k$, the generalized entropy becomes
\be
S_{\rm gen} =\f c 6\left(-x_1^+ x_1^-   -x_2^+ x_2^-  +2 k +  \log \left(\f{(x_2^+-x_1^+)(x_1^- - x_2^-)}{\epsilon_{\rm rg}^2}\right)\right).
\ee
The quantum normal region $\pm \partial_{x_2^\pm} S_{\rm gen} \geq 0$ for $x_1^{\pm} = x_2^{\mp}$ gives 
\be
x_2^+ \leq x_2^- + \f{1}{x_2^-}\,.
\ee
The above condition comes from the outgoing constraint, as the ingoing one is strictly weaker. The quantum normal region is illustrated in figure \ref{fig:eternalBH}, and the quantum extremal surface is seen to lie within it as required. 

The quantum normal region for $G$ is computed similarly, where $G$ is the union of the interval from $(x_2, t_2)$ to $(x_G, t_G)$ and its reflection $x \rightarrow -x$. We will assume the entropy factorizes into the sum of entropies of the two intervals. Extremizing with respect to the left endpoint of the right interval $\partial_{x_2^{\pm}} S_{\rm gen} = 0$ gives 
\be\label{extrem}
x_2^\pm = \f{1}{x_2^\mp-x_G^\mp}
\ee
In the limit $x_G^+\rightarrow \infty$, we find the quantum normal region 
\be
x_2^- \geq 0\,,\qquad x_2^+ \leq \f{1}{x_2^- - x_G^-}\,.
\ee
Thus the quantum extremal surface must be on or inside the horizon. The limit $x_G^+ \rightarrow \infty$ restricts region $R$ to be on $\mathcal{I}^+$ (and lie at times $x^- < x_G^-$), in which case the island was found to lie on the horizon. Notice that in the limit $x_G^+\rightarrow \infty$ the extremizations done to obtain \eqref{extrem} are precisely the extremizations necessary to find an island for a region $R = (I \cup G)^c$. This means that the endpoint of the island is given as $x_2^+ = -1/x_G^-$, $x_2^- = 0$, consistent with \cite{Hartman:2020swn}. In the limit $x_G^- \rightarrow -\infty$, region $R$ has to vanish and the quantum normal region for $G$ shrinks to the bifurcation point, consistently reproducing the classical extremal surface.

Our final example is provided by an evaporating black hole in the same model. This solution has a shock wave impinging on the vacuum $T_{\s^\pm \s^\pm} = 0$ where $\s^\pm = \pm \log \pm x^\pm$. The region $R$ and its island $I$ are shown in figure \ref{fig:evap-regions}. At late times and large initial mass, the regulated matter entropy of region $I$ is given by 
\be
\widehat{S}_{\rm mat} \approx \f{c}{12}\log x^+_{QES} \approx -\f{c}{12}\log \left(-x_2^-\right)\,,
\ee
while the gravitational entropy is given by 
\be
S_{\rm grav} = \f{c}{24} (4M +\log (-x_2^-))
\ee
Thus for $\log(-x_2^-) = -4M/3$ the Bekenstein area bound is violated. This is precisely the Page transition. 

We now compute the quantum normal region for this solution.  The entanglement entropy of region $I$ (for consistency with the previous example, we will refer to the right endpoint of $I$ as $(x_2^+, x_2^-)$, even though there is no left endpoint) is given by
\be
\f c 6 \left(\rho + \f 1 2 \log (-x_2^+ x_2^-) + \log \log (-4 x_2^+ x_2^-)\right)
\ee
We once again use $\rho + 6 S_{\rm grav}/c = \Omega + k$ where $\Omega = -x^+ x^- -\f 1 4 \log(-4 x^+ x^-) -M(x^+-1)\Theta(x^+-1)$ for the evaporating black hole and we consider $x_2^+ > 1$. The generalized entropy is therefore
\be
S_{\rm gen}= \f c 6 \left(\f 1 2 \log (-x_2^+ x_2^-) + \log \log (-4 x_2^+ x_2^-) - x_2^+ x_2^- - \f 1 4 \log(-4 x_2^+ x_2^-) - Mx_2^+ + k\right)
\ee
where $M$ is a parameter related to the strength of the shock wave and therefore the mass of the resulting black hole. The quantum extremal surface lies on the curve \cite{Hartman:2020swn}
\be\label{rstcurve}
4(M+ x^-_{2})x_{2}^+ - 1 =0 \ .
\ee
The quantum normal region is defined by $\pm \partial_{x_2^\pm} S_{\rm gen}\geq 0$, which requires
\be\label{qesqnm}
 4(M+ x_2^-) x_2^+ - 1 - \f{4}{\log(-4x_2^+ x_2^-)} \leq 0\,.
\ee
Since $x_2^-<0$, we see that the quantum extremal surface is inside the quantum normal region. 

%The equation for the quantum extremal surface is given in \cite{Hartman:2020swn} and the quantum normal region is given by $\pm \partial_{x_2^\pm} S_{\rm gen}\geq 0$. The outgoing quantum normal condition is strictly stronger than the ingoing one, and altogether we have
%\be\label{qesqnm}
%4(M+ x^-_{QES})x^+ - 1 =0\,,\qquad 4(M+ x^-) x^+ - 1 - \f{4}{\log(-4x^+ x^-)} \leq 0\,.
%\ee
%Since $x^-<0$, we see that the quantum extremal surface is inside the quantum normal region. 

We would like to consider the quantum normal region for $G$. Instead of taking the right endpoint $x_G$ to be on $\mathcal{I}^+$ as for the eternal black hole, we instead place it inside the black hole at the evaporation endpoint, where the apparent horizon meets the singularity. Thinking of the black hole interior as a baby universe, this is the analog of picking $G$ to be $I^c$ in the cosmological half of the thermofield double. 

The evaporation endpoint is given by 
\be
x_G^+ = \f{1}{4M}\left(e^{4M}-1\right)\,,\qquad x_G^- = \f{M}{e^{-4M}-1}\,.
\ee
The generalized entropy is given in section 3.2 of \cite{Hartman:2020swn}, resulting in a quantum normal region 
\be
x_2^+ \leq \f{1}{4x_2^-}\left(1+\f{4}{\log \f{x_2^-(e^{-4M}-1)}{M}}\right)\,,\qquad x_2^- \geq -M+\f{1}{4x_2^+}-\f{1}{x_2^+ \log \f{e^{4M}-1}{4Mx_2^+}}\,.
\ee
For $M \gg 1$ and $e^{4M} \geq x_2^+ \gtrsim  e^{\mathcal{O}(M)}$, the latter inequality saturates at the actual location of the QES, as seen by explicit comparison to \eqref{rstcurve}. Thus the QES sits on the border of the allowed region.

\section{Derivation of the bound on $|\p \Ssemi|$}\label{app:fderiv}

In this appendix we review the derivation of the second inequality in \eqref{fderivs} bounding spacelike derivatives of the matter entropy density, following \cite{1512.02695}. See also \cite{1509.05044}.

Let $\rho$ be the density matrix of the matter fields in the thermofield double, and $\rho_A$ its reduction to a region $A$. The state $\rho_{\rm th} = \rho_I \otimes \rho_R$ can be thought of as the density matrix of two copies of a thermal state reduced to $I \cup R$. Unlike the thermofield double, this state has no entanglement between $I$ and $R$. Up to sub-extensive corrections, the relative entropy of $\rho_{I\cup R}$ with respect to this state is given by \cite{1512.02695}
\be
S_{\rm rel}(\rho_{I \cup R} | \rho_I \otimes \rho_R ) \approx s_{\rm th} (V_I + V_R) - \Sred(I \cup R) \ .
\ee
Monotonicity of relative entropy requires 
\be
\p_{r_I} S_{\rm rel}(\rho_{I \cup R} | \rho_I \otimes \rho_R ) \geq 0 \ , 
\ee
which implies $\p_{r_I} f \leq 1$. Another way to reach the same result is to note that this relative entropy is equal to the mutual information $I(I,R)$, so monotonicity is equivalent to strong subadditivity.

To bound $\p_{r_I} f$ from below, we use strong subadditivity, $S_X + S_Y \geq S_{X \cup Y} + S_{X \cap Y}$. Let $I(r)$ denote the region of size $r$, so $I = I(r_I)$, and choose $X = I(r_I+\delta r) \cup R$, $Y = I(r_I) \cup I(r_I + \delta r + \gamma)^\setcomp$, where the complement includes a potential purifying system. Here $\delta r$ is the small deformation corresponding to the $\p_{r_I}$ derivative and $\gamma \ll \delta r$ is a geometric regulator similar to the strip of size $\delta$ in figure \ref{fig:ssa-regions-2}. The UV divergences cancel in SSA. Taking $\gamma$ small and keeping only the extensive contributions, SSA becomes 
\be
\Sred(I(r_I + \delta r)\cup R) \gtrsim \Sred(I(r_I)\cup R) - \Sred(I(r_I + \delta r) \backslash I(r_I)) \ .
\ee
Here `$\gtrsim$' indicates that we keep only the extensive parts of $\Sred$.
This requires $\p_{r_I} f > - 1$, so together with the result above we have derived $|\p_{r_I}f| \leq 1$.

\renewcommand{\baselinestretch}{1}\small
\bibliographystyle{ourbst}
%\bibliography{replicaBib}
\bibliography{frw-draft}
\end{document}